\documentclass[a4paper,preprint,12pt,sort&compress]{elsarticle}

\usepackage{amsmath,amssymb,amsthm,color,xfrac,mathrsfs}
\usepackage{bbm}
\usepackage{makecell}
\usepackage{epsfig,amsfonts}
\usepackage{graphicx,epsfig,epstopdf} 
\usepackage{array} 
\usepackage{color,soul}
\soulregister\cite{7}
\soulregister\citep{7}
\soulregister\ref{7}
\soulregister\pageref{7}
\usepackage[usenames,dvipsnames]{xcolor}
\sethlcolor{yellow}
\usepackage{appendix}

\usepackage{indentfirst} 
\usepackage{enumerate} 
\usepackage{lineno}
 \usepackage{placeins} 
\usepackage{booktabs,ctable,multirow,longtable} 
 \usepackage{rotating}

\usepackage{changepage}

\DeclareMathAlphabet{\mathpzc}{OT1}{pzc}{m}{it}
\DeclareMathAlphabet{\mathcalligra}{OT1}{calligra}{m}{it}
\newcommand{\eqn}[2]{\begin{equation} \label{#1} {#2} \end{equation}}
\newcommand{\eqnn}[1]{\begin{equation*}  {#1} \end{equation*}}

\newcommand{\bs}[1]{\boldsymbol{#1}}
\newcommand{\pfw}{\mathpzc{w}(d)}
\newcommand{\pfwP}{\mathpzc{w}^\prime(d)}
\newcommand{\fpsi}{{\overline \alpha}}

\AtBeginDocument{
\let\div\undefined\DeclareMathOperator{\div}{{div}}
}

\newcolumntype{P}[1]{>{\centering\arraybackslash}m{#1}}

\graphicspath{{./Images/EPS/}}

\newtheorem{remark}{Remark}
\journal{arXiv}

\begin{document}
\begin{frontmatter}


\title{A novel framework to model the fatigue behavior of brittle materials based on a variational phase-field approach}
\author[1]{P. Carrara}\corref{cor1}
\ead{p.carrara@tu-braunschweig.de}
\author[1]{M. Ambati}
\author[2]{R. Alessi}
\author[1]{L. De Lorenzis}

\address[1]{Technische Universit{\"a}t Braunschweig, Institute of Applied Mechanics, Pockelsstr. 3, 38106 Braunschweig, Germany}
\address[2]{Universit{\`a} di Pisa, Department of Civil and Industrial Engineering, Largo Lucio Lazzarino, 56122 Pisa, Italy}
\cortext[cor1]{Corresponding author}

\begin{abstract}
A novel variational framework to model the fatigue behavior of brittle materials based on a phase-field approach to fracture is presented. The standard regularized free energy functional is modified introducing a fatigue degradation function that effectively reduces the fracture toughness as a proper history variable accumulates. This macroscopic approach allows to reproduce the main known features of fatigue crack growth in brittle materials. Numerical experiments show that the W{\"o}hler curve, the crack growth rate curve and the Paris law are naturally recovered, while the approximate Palmgren-Miner criterion and the monotonic loading condition are obtained as special cases.
\end{abstract}

\end{frontmatter}


\section{Introduction and state of the art} \label{sct:intro}

	The term fatigue refers to repeatedly applied macroscopic loads or displacements whose maximum value is below the monotonic strength of the material \citep{Carrara2015}. When a component is subjected to fatigue loading (cyclic or not), it experiences, at first, the formation of microdefects (e.g., microvoids) at micro-heterogeneities such as pits or imperfections.  Depending on the type of material, within this phase energetic barriers that inhibit the microdefects evolution when the load level is below a so-called fatigue threshold might be present \citep{Newman1998,Schijve2003}. When the fatigue process advances, the microdefects evolve into microcracks. These early processes are ruled by the stochastic micro-structural arrangement of the material, hence they are random in nature. The microcracks eventually coalesce and lead to the formation of a fatigue (macro-)crack, whose size is sufficient to neglect the aleatory nature of the material microstructure. This macro-crack then propagates first stably and finally unstably leading to failure.

\begin{figure}[!h]
	\begin{adjustwidth}{-3cm}{-3cm}
	\centering
		\includegraphics{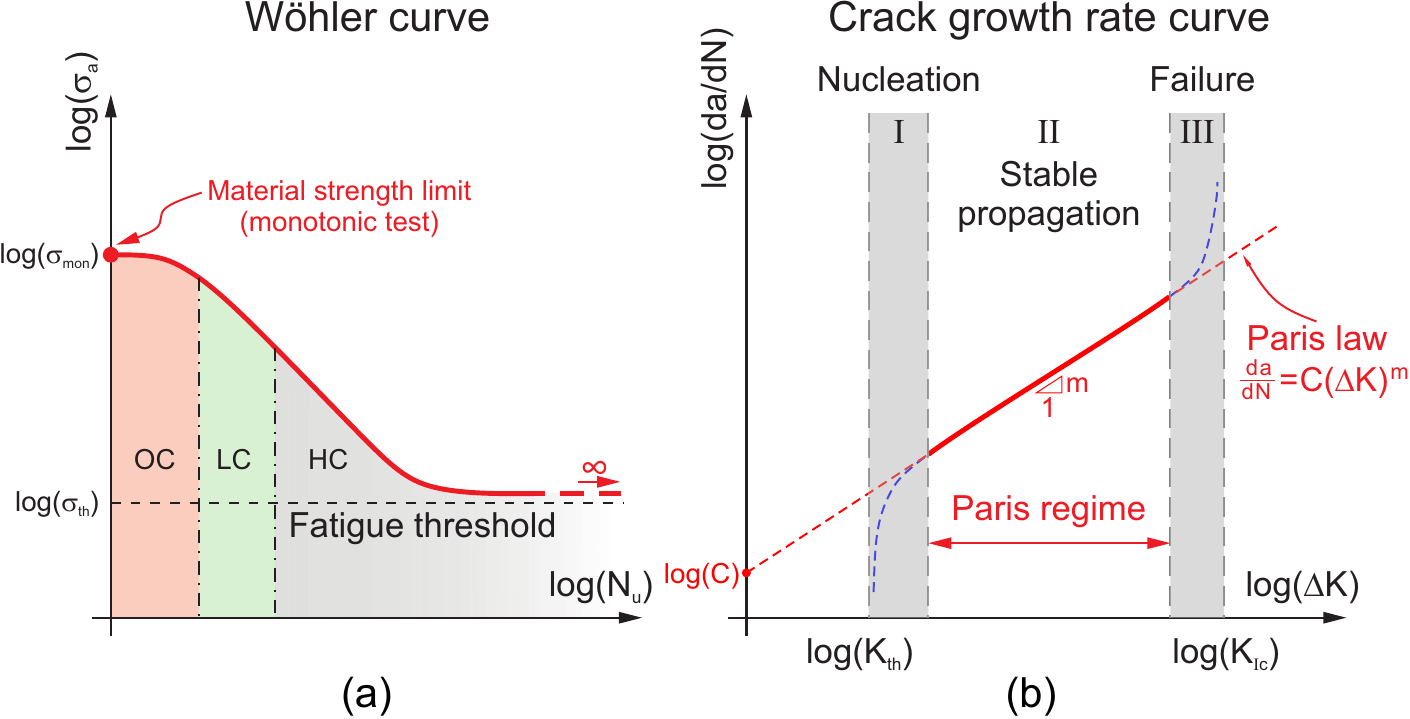}                                                                                             
	\end{adjustwidth}
		\caption{(a) W{\"o}hler or $S-N$ curve. (b) Crack growth rate curve and Paris law.} 
		\label{fig:fatigue_mod}
	\end{figure}

	The early studies on fatigue are mostly empirical and based on the data fitting of vast experimental campaigns \citep{Suresh1998}. In \citep{Wohler1870} W{\"o}hler (1870) studies fatigue using experimental curves relating the maximum number of cycles that a component can undergo before failure, $N_u$, to the (constant) applied stress amplitude $\sigma_a$ (Fig.~\ref{fig:fatigue_mod}a). This curve, named W{\"o}hler or $S-N$ curve, is still used and is mathematically formalized by, e.g., the Basquin relationship
	
	\eqn{eq:basquin}{\sigma_a=AN_u^\beta\,,}
	
	\noindent where $A$ and $\beta$ are empirical coefficients that depend on the geometry and test setup. W{\"o}hler's approach catches some characteristic features of fatigue such as the leading role of the load amplitude, the presence of an upper stress amplitude related to the monotonic strength of the material and the (possible) presence of a fatigue threshold. We divide here the W{\"o}hler curve in three regions corresponding to (i) oligocyclic (\textsf{OC}), (ii) low-cycle (\textsf{LC}) and (iii) high-cycle (\textsf{HC}) fatigue (Fig.~\ref{fig:fatigue_mod}a). The \textsf{OC} region is related to high load levels, therefore $N_u$ is rather limited, the material is largely damaged already after the first cycles and the stable crack propagation phase tends to disappear. In the \textsf{LC} region damage and fatigue processes compete, leading to a stable crack propagation phase whose extension is comparable to the nucleation and failure phases. In the \textsf{HC} region the  fatigue life is dominated by the stable crack propagation phase. If the material features a fatigue threshold, this region asymptotically approaches an infinite fatigue life branch. 
		
	Palmgren (1924) \citep{Palmgren1924} and Miner (1945) \citep{Miner1945} introduce for the first time the concept of cumulative damage in the study of fatigue. Following the work of W{\"o}hler, they postulate that, if there are $k$ different stress amplitudes in a loading history, one cycle performed at $\sigma_a^i$ ($1 \le i \le k$) gives a contribution to the overall damage equal to $\gamma_i=1/N_u(\sigma_a^i)$ independently on the sequence of cycles. The following failure criterion\footnote{Note that in the original contribution \citep{Miner1945} failure is intended as the onset of a crack.} is then proposed
	
	\eqn{eq:miner}{\sum_{i=1}^{k} n_{\sigma_{a}^i}\gamma_i= \sum_{i=1}^{k}\frac{n_{\sigma_{a}^i}}{N_u(\sigma_{a}^i)}=1\,.}

\noindent where $n_{\sigma_{a}^i}$ is the total number of cycles done at $\sigma_{a}^i$. 
 
	The formalization of the fracture mechanics theory of Griffith (1921) \citep{Griffith1921} radically changes the study of fatigue. Paris (1961)  \citep{Paris1961}, after the work of Irwin (1957) \citep{Irwin1957}, has the pioneering idea of proposing the stress intensity factor range in a single fatigue cycle $\Delta K$ as driving force for the fatigue growth of a crack with length $a$. The stress intensity factor summarizes locally at the crack tip the influences of the external loads, boundary conditions and geometry, making local an apparently structural problem. Fig.~\ref{fig:fatigue_mod}b shows an illustrative crack growth rate curve $d a/dN$ vs. $\Delta K$ in a bi-logarithmic scale obtained from constant amplitude cyclic fatigue tests. In general, this curve permits to distinguish three different regions related respectively to the crack nucleation ($I$), stable propagation ($II$) and unstable propagation ($III$). The extension of the stable propagation branch can vary depending on the severity of the applied load (i.e., \textsf{OC}, \textsf{LC} or \textsf{HC} regimes in Fig.~\ref{fig:fatigue_mod}a). Due to the microstructural-related nature of crack nucleation, the nucleation branch is usually affected by a high scatter. However, when a singularity in the component is present, e.g., a pre-existing notch, the scatter decreases and the crack nucleation and successive propagation can be assumed as nearly-deterministic.

	Paris and Erdogan (1963) \citep{Paris1963} proposed the following relationship, known as Paris law, to describe the stable propagation of a fatigue crack (red branch of the curve in Fig.~\ref{fig:fatigue_mod}b)
	
	\eqn{eq:paris}{\frac{d a}{d N} = C\Delta K^m\,.}
	
\noindent The two constants $C$ and $m$ need to be experimentally calibrated and are meant to be material parameters. Eq.~\ref{eq:paris} is valid only for low crack growth rate (i.e., $d a/d N\le$10$^{-4}$mm/cycle following \cite{ASTM_647,ASTM_1820}), within the so-called Paris regime and is unable to reproduce the crack nucleation and failure phases (Fig.~\ref{fig:fatigue_mod}b). Trying to overcome its limits, Eq.~\ref{eq:paris} has been constantly improved and extended until reaching, today, the form of the widespread \emph{NASGRO} equation \citep{NASGRO}, that reproduces many characteristic aspects of the fatigue behavior, such as the nucleation, propagation and failure phases, the crack closure effect \citep{Elber1970}, the presence of different cracking modes and the effect of the maximum load reached within the cycle \citep{Rabold2014} but at the cost of introducing up to 11 parameters \citep{Rabold2013}.

	The principal drawbacks related to the early approaches such as the W{\"o}hler and Palmgren-Miner rules (and their extensions) are rooted in their empirical nature, that makes them hardly extendable to conditions different from the specific situation tested. Moreover, they focus only on failure and not on crack growth. Also, the W{\"o}hler curve is limited to the cases of constant amplitude cycles. The Palmgren-Miner rule is valid only when the order of application of the cycles does not influence the results. This happens in some special cases reported in \ref{app:miner_appr} or when the applied load is random. Conversely, the approaches based on Paris law are limited by the necessity of knowing $\Delta K$ which is, in general, a function of the crack length, geometry and boundary conditions. Analytical relationships exist only for few cases with very simple geometries and boundary conditions. In more general cases they can be found with numerical approaches such as the finite element method (FEM). 

	The numerical simulation of fatigue crack growth is much more flexible than the use of analytical methods but involves a double-fold challenge: on one hand we need a suitable way to represent the crack and, on the other hand, the latter should evolve  as a result of fatigue for loads below the monotonic strength. The former issue is common to the monotonic case and the advantages and shortcomings of the available approaches are summarized, e.g., in \citep{Ambati2014}. Concerning the latter, here we limit ourselves to highlight the major pros and cons of each class of methods, while extensive reviews can be found in \citep{Rege2017,Branco2015,Desmorat2006}.  

   The FEM or the extended FEM (XFEM) can be used to numerically obtain the $\Delta K$ values to be used in Paris-like laws \citep{Rabold2013,Rege2017,Branco2015}. For some standard geometries, parameterized relationships based on best fitting procedures are also available \citep{Tada2000} and adopted by material testing guidelines such as \cite{ASTM_647, ASTM_1820}. This approach is limited by the necessity of assuming an initial notch and needs criteria to determine the direction and shape evolution of the crack. Even more critical is the need to update the geometry of the problem when the crack front advances, which becomes extremely complex in 3-D \citep{Rege2017}. This procedure is also hardly applicable to multiple cracks especially in case of merging/branching phenomena, complex geometries and/or loading conditions. Another family of approaches, including e.g. crystal plasticity and molecular dynamics models \citep{Sha2017,Kozinov2018,Li2017}, describes the material behavior at a very small scale where the microstructure cannot be neglected. While these approaches contribute an important insight into the physics of the fatigue crack nucleation phenomena, their adoption for domains representative of real scale components is at the present stage unfeasible. Rather, they can be used to investigate the uncertainty in the fatigue crack nucleation phase as a result of the distribution of microstructural imperfections.  

   An issue common to all approaches is the calibration of the parameters. All the methods based on linear elastic analyses to obtain the $\Delta K$ values and many continuum damage or cohesive interface models use a Paris-like law and/or the W{\"o}hler curve as input rather than obtaining them as output \citep{Zi2004,Rege2017,Branco2015}. Other models are deemed to reproduce only a specific fatigue-related aspect disregarding the general behavior \citep{Salvati2017,Kozinov2018}. Another limitation of some continuum damage and plasticity approaches \citep{Desmorat2006,Hosseini2018,Holopainen2016} is the definition of evolution equations for the fatigue process that are often uncorrelated with the fracture mechanics or involve a large number of parameters with unclear physical meaning.

	The variational phase-field approach to fracture \citep{Francfort1998,Bourdin2000} is very attractive to model crack nucleation and propagation. It describes a steep but smooth transition from intact to fully cracked material states by means of an order parameter termed phase-field variable \citep{Ambati2014}. The approach can recover Griffith's theory as a limit case in the $\Gamma$-convergence sense, and at the same time can be classified as a gradient damage approach \citep{Pham2011}. The framework is very attractive because it can easily deal with complex crack patterns in 3-D with no need for remeshing nor for particular criteria to track the crack propagation. Boldrini et al. \citep{Boldrini2016} proposed a phase-field model that couples the cracking behavior with the thermal and fatigue problem. The fatigue effect is introduced as an additional order parameter and its evolution is postulated under some restrictive conditions to preserve the thermodynamic consistency. Also, in \citep{Caputo2015,Amendola2016} the authors adopt the Ginzburg-Landau formalism to formulate a phase-field model accounting for fracture, visco-elasticity and environmental effects. Here, a fatigue potential is introduced to allow the degradation of the material under fatigue loadings. In both cases, no evidence is given that the proposed model reproduces the major features of the fatigue behavior.

  In general, a framework that is able to reproduce both the mechanics of monotonic fracture (comprising nucleation, propagation and failure) and the known features of the fatigue behavior including the Paris law, the W{\"o}hler curve with the transition between oligo-, low- and high-cyclic fatigue and the Palmgren-Miner law is still missing.		

	In the present paper we propose a novel approach to model the fatigue behavior of brittle materials based on variational phase-field modeling of fracture. The first 1-D investigation has been reported in \citep{Alessi2017} and the existence of quasi-static evolutions with a vanishing viscosity approach studied in \citep{Alessi2018}. As in \citep{Alessi2017} we modify the free energy potential of the monotonic case introducing a suitable fatigue history variable and a fatigue degradation function that modifies the rate of the dissipated energy accounting for the fatigue loading history. The proposed approach aims at linking regularized monotonic fracture mechanics to fatigue crack growth establishing a framework suitable for any type of (brittle) materials. We are able to reproduce the major features of the fatigue behavior including the crack nucleation, stable and unstable propagation phases. Also, the Paris law and the W{\"o}hler curve are recovered naturally, while the Palmgren-Miner rule and the monotonic behavior are encompassed as special cases.
	  
	  The paper is structured as follows: the adopted phase-field model of brittle fracture is briefly summarized in Section~\ref{sct:monotonic} and extended to fatigue in Section~\ref{sct:fatigue}. Sections~\ref{sct:comp_asp} and \ref{sct:num_ex} illustrate respectively the details of the numerical implementation and the numerical examples, while conclusions are drawn in Section~\ref{sct:concl}.

\section{Starting point: monotonic loading} \label{sct:monotonic}
In this section we briefly recall the phase-field model of brittle fracture under monotonic loading adopted as starting point \citep{Bourdin2000, Ambati2014,Miehe2010,Pham2011}. Isothermal conditions, negligible inertial effects and smooth loading in time are assumed. This allows to rely on the energetic principles of rate-independent systems \citep{Mielke2015}, in the form of an \emph{energy balance}, a \emph{dissipation inequality} and a \emph{stabilty criterion} applied to a properly defined set of energetic quantities. Also, the assumptions of small strains and of irreversibility for any dissipative process are assumed to hold. 

\subsection{Phase-field modeling of brittle fracture under monotonic loading} \label{sct:PF}
Consider a linear elastic  $D$-dimensional body $\Omega$ susceptible of brittle fracture. The internal energy density is assumed as

\eqn{eq:energy}{W\left(\bs{\varepsilon}(\bs{u}),d,\nabla d\right)=\psi_{el}\left(\bs{\varepsilon}(\bs{u}), g(d)\right) + \varphi \left(d,\nabla d \right)\,,}

\noindent where $\psi_{el}\left(\bs{\varepsilon}(\bs{u}), g(d)\right)$ is the stored elastic energy density and $\varphi \left(d,\nabla d \right)$ is the fracture energy density. Also, $\bs{\varepsilon}$ is the infinitesimal strain tensor related to the displacement field $\bs{u}$ by $\bs{\varepsilon}=\nabla^s (\bs{u})$, $\nabla^s$ being the symmetric gradient operator, while $d$ is the scalar phase-field parameter varying smoothly from 0 (sound material) to 1 (broken material). The \emph{degradation function} $g(d)$ governs the transition of the mechanical behavior of the material from the sound to the cracked state, with

\eqn{eq:a_prop}{g(0)=1\,, \quad g(1) = 0\,, \quad g^\prime(d) \le 0 \,\text{  for  }\, 0\le d\le 1\,.}

 The fracture energy density is written as \citep{Ambati2014,Pham2011,Bourdin2000}

\eqn{eq:pf_fract_en}{\varphi \left(d,\nabla d \right)=\frac{G_c}{4 c_w}\left(\frac{\pfw}{\ell}+\ell|\nabla d|^2\right)\quad \text{with} \quad c_w=\int_0^1\sqrt{\mathpzc{w}(\delta)} \, d\delta \,,}

\noindent where $\ell$ is a regularization length, $G_c$ is the fracture toughness of the material and $\pfw$ is termed \emph{dissipation function}. In general, $\pfw$ must fulfill the following properties \citep{Braides1998}

\eqn{eq:w_prop}{\mathpzc{w}(0)=0\,, \quad \mathpzc{w}(1) = 1\,, \quad \mathpzc{w}^\prime(d) \ge 0 \,\text{  for  }\, 0\le d\le 1\,.}

  Substituting Eq.~\ref{eq:pf_fract_en} into Eq.~\ref{eq:energy} and integrating over $\Omega$ gives the free energy functional 

\eqn{eq:free_en}{E_{\ell}(\bs{u},d
)=\int_\Omega\psi_{el}\left(\bs{\varepsilon}(\bs{u}), g(d)\right)d\bs{x} + \frac{G_c}{4 c_w}\int_\Omega\left(\frac{\pfw}{\ell}+\ell|\nabla d|^2\right)d\bs{x}\,.}

\noindent  $\Gamma$-convergence results ensure that the global minima of $E_{\ell}$ in Eq.~\ref{eq:free_en} converge to those of the unregularized functional as $\ell \rightarrow 0$ \citep{Braides1998,Francfort1998,Ambrosio1992}.

Applying the energy principles to Eq.~\ref{eq:free_en}, along with the irreversibility condition

\eqn{eq:irrev}{\dot d\ge 0\,,}

\noindent leads to the governing equations of the problem in terms of momentum balance and crack propagation conditions and respective boundary conditions (see sect.~\ref{sct:fatigue}).

\subsection{Degradation function and dissipation function} \label{sct:diss_fct}

The degradation function describes the smooth degradation of the material behavior from sound to fully cracked state. Although alternatives are proposed in the literature \citep{Alessi2015,Wilson2013,Steinke2018}, the present work is limited to the analysis of the well known relationship \citep{Ambati2014,Bourdin2000,Miehe2010}

\eqn{eq:degradation}{g(d)=(1-d)^2\,.}

The dissipation function rules the energy dissipation due to the formation of a new crack (Eq.~\ref{eq:pf_fract_en}). Here we adopt two widely used models \citep{Tanne2018}

\begin{enumerate}

\item \texttt{AT2} model: originally proposed by Ambrosio and Tortorelli \citep{Ambrosio1992} and then adopted in \citep{Bourdin2000} and many following works

\eqn{eq:w_AT2}{\pfw=d^2 \quad \text{  and  }\quad c_w=\frac{1}{2}\,.}

\noindent As $\pfwP=0$, this model features a vanishing threshold for the onset of damage, leading to a material model without an initial linear elastic branch. 

\item \texttt{AT1} model: proposed in \citep{Pham2011} with the aim of reproducing a constitutive  behavior with an initial linear elastic branch. It reads

\eqn{eq:w_AT1}{\pfw=d\quad \text{  and  }\quad c_w=\frac{2}{3}\,.}

\end{enumerate}

\subsection{Decomposition of the elastic strain energy density} \label{sct:el_en}
 An additive decomposition of the undamaged elastic strain energy density into active and inactive parts is needed to describe the tension/compression asymmetry in the material behavior. The elastic strain energy density $\psi_{el}\left(\bs{\varepsilon},g(d)\right)$ is assumed as 

\eqn{eq:deg_en}{\psi_{el}\left(\bs{\varepsilon},g(d)\right)=g(d)\psi^+_{el,0}(\bs{\varepsilon})+\psi^-_{el,0}(\bs{\varepsilon})=\psi_{el}^+(\bs{\varepsilon},d)+\psi^-_{el}(\bs{\varepsilon})\,,}

\noindent where $\psi^+_{el,0}(\bs{\varepsilon})$ and $\psi_{el,0}^-(\bs{\varepsilon})$ are the active and inactive parts of the undamaged elastic strain energy density $\psi_{el,0}$. For an isotropic material $\psi_{el,0}=\psi_{el,0}^++\psi_{el,0}^-=\frac{1}{2} \lambda \, tr^2(\bs{\varepsilon}) + \mu \, tr (\bs{\varepsilon})^2$, $\lambda$ and $\mu$ being the Lam{\'e} constants. Accordingly, the stored elastic strain energy density of a damaged material is split into an active part $\psi_{el}^+(\bs{\varepsilon},g(d))$ that is degraded and an inactive part $\psi_{el}^-(\bs{\varepsilon})$ not affected by the phase field parameter.  Although other options are available \citep{Steinke2018,Lancioni2009,Strobl2016}, the present work adopts the following choices:

\begin{enumerate}
\item \texttt{Isotropic model}: proposed in \citep{Bourdin2000}, it features the degradation of the whole elastic strain energy density. Hence, it lets the fracture propagate also in compressed regions \citep{Amor2009}. 

\item \texttt{Volumetric/deviatoric split}: proposed in \citep{Amor2009} to overcome the drawbacks of the isotropic model. The degradation function affects only the energy density related to the deviatoric and to the positive volumetric part of the strain tensor. 

\item \texttt{Spectral split}: proposed in \citep{Miehe2010}, it distinguishes between degraded and undegraded parts of the energy using the spectral decomposition of the strain tensor.

\item \texttt{No-tension split}: proposed in \citep{Freddi2010}, it degrades the energy related to the positive-definite symmetric part of the strain tensor leaving undegraded the remaining part.

\end{enumerate}

\subsection{Homogeneous 1-D solution of the phase-field problem} \label{sct:hom_sol}
Studying analytically the homogeneous 1-D solution \citep{Pham2011}, it is possible to compute the peak stress $\sigma_y$ and the corresponding strain $\varepsilon_y$ as functions of the Young's modulus $E$, the regularization length $\ell$  and the fracture toughness $G_c$ as follows

\eqn{eq:hom_AT2}{\sigma_y^{AT2}=\frac{9}{16}\sqrt{\frac{E G_c}{3\ell}}\,,\quad\quad \varepsilon_y^{AT2}=\sqrt{\frac{G_c}{3\ell E}}\,,}

\noindent for the \texttt{AT2} model and

\eqn{eq:hom_AT1}{\sigma_y^{AT1}=\sqrt{\frac{3E G_c}{8\ell}}\,,\quad\quad \varepsilon_y^{AT1}=\sqrt{\frac{3G_c}{8\ell E}}\,,}

\noindent for the \texttt{AT1} model.


\section{Extension to fatigue behavior} \label{sct:fatigue}
In this section the phase-field modeling approach to brittle fracture in section~\ref{sct:monotonic} is extended to fatigue. 

		\subsection{Energetic quantities} \label{sct:en_quantities}
To introduce the fatigue effects, we propose to modify the fracture energy density similarly to \citep{Alessi2017} as follows

\eqn{eq:fatigue_surf}{\begin{split}\varphi_F \left(d,\nabla d | \fpsi ([0,t]) \right)&=\int_0^t f(\fpsi(\tau)) \dot \varphi(d,\nabla d)d\tau\\
&=\int_0^t  f(\fpsi(\tau)) \left(\frac{\partial \varphi}{\partial d}\dot d + \frac{\partial \varphi}{\partial \nabla d} \cdot \nabla \dot d\right) d\tau\,, \end{split}} 

\noindent where $t$ is the pseudo-time, $\fpsi(t)$ is a properly defined cumulated history variable acting in Eq.~\ref{eq:fatigue_surf} as a parameter, while  $f(\fpsi(t))$ is a \emph{fatigue degradation function}. 

\begin{remark}[Dissipated energy]
In this framework the dissipated energy is a process dependent quantity and no longer a state function.
\end{remark}

\begin{remark}[Choice of $\fpsi(t)$]
The history variable $\fpsi(t)$ is not yet defined at this stage. It can be taken as a cumulation of any scalar quantity $\alpha$ which can exhaustively describe the fatigue history experienced by the material so that
\eqn{eq:irr_state}{\dot \fpsi(t) = |\dot \alpha| \ge 0\,.}

\end{remark}

\begin{remark}[Dissipation]
The dissipation is still related only to the damage variable, because $\fpsi(t)$ acts in Eq.~\ref{eq:fatigue_surf} merely as a parameter that tunes the dissipation rate accounting for the load history experienced by the material. In other words, while the effective dissipative work is still due to the evolution of the phase-field variable and its gradient, the function $f(\fpsi(t))$ effectively modulates the fracture toughness as a function of the ``mileage'' as expressed by the variable $\fpsi(t)$.  
\end{remark}

\begin{remark}[Support of the phase field]
Substituting Eq.~\ref{eq:pf_fract_en} into Eq.~\ref{eq:fatigue_surf} it is

\eqn{eq:diss_rateF}{\dot\varphi_F \left(d,\nabla d | \fpsi ([0,t]) \right)=f(\fpsi(t)) \frac{G_c}{4 c_w}\left(\frac{\pfwP}{\ell}\dot d+2\ell \,\nabla d \cdot \nabla \dot d\right)\,,}

\noindent where it is evident that the function $f(\fpsi(t))$ affects both local and gradient terms of the dissipative power. This choice ensures that the support of the phase-field variable remains the same as in the monotonic case \citep{Miehe2010}.
\end{remark}

The function $f(\fpsi(t))$ is assumed to have the following properties

\eqn{eq:d_prop}{f(\fpsi\le\alpha_T)=1\,, \quad f(\fpsi>\alpha_T) \in [0,1] \,, \quad f^\prime(\fpsi) \le 0 \,\text{  for  }\, 0\le f(\fpsi) < 1\,,}

\noindent where $\alpha_T$ is a threshold controlling when the fatigue effect is triggered. 

The total internal energy density assumes now the form

\eqn{eq:energyF}{\begin{split}W &\left(\bs{\varepsilon}(\bs{u}),d,\nabla d | \fpsi\right)=\psi_{el}\left(\bs{\varepsilon}(\bs{u}), g(d)\right) +\\&+ \int_0^t  f(\fpsi(\tau)) \frac{G_c}{4 c_w}\left(\frac{\pfwP}{\ell}\dot d+2\ell\, \nabla d \cdot \nabla \dot d\right) d\tau\,,\end{split}}

\noindent thus becoming (time-)history-dependent. To circumvent this dependency the energetic principles are here applied to the total internal power density 

\eqn{eq:powerF}{\begin{split}\dot W\left(\bs{\varepsilon}(\bs{u}),d,\nabla d | \fpsi\right)=&\,\bs{\sigma}:\bs{\dot \varepsilon}(\bs{u})+\frac{\partial \psi_{el}}{\partial d}\dot d+\\&+  f(\fpsi(t)) \frac{G_c}{4 c_w}\left(\frac{\pfwP}{\ell}\dot d+2\ell \, \nabla d \cdot \nabla \dot d\right)\,.\end{split}}

\noindent where the Cauchy stress tensor $\bs{\sigma}$ is introduced as

\eqn{eq:Cauchy}{\displaystyle \bs{\sigma}=\frac{\partial \psi_{el}}{\partial \bs{\varepsilon}}\,.}

\subsection{Governing equations} \label{sct:gov_eqs}

Applying now the energetic principles for a rate-independent system in rate form it is possible to determine the governing equations of the problem. The solution $(\bs{ u}, d)$ satisfies the energy balance principle if the following condition holds

\eqn{eq:eb}{\int_\Omega \dot W\left(\bs{\varepsilon}(\bs{u}),d,\nabla d | \fpsi\right)d\bs{x}- \mathcal{\dot L(\bs{u})} =0\,.}

\noindent Here $\mathcal{\dot L}(\bs{u})$ is the external power, given by

\eqn{eq:ext_power}{\mathcal{\dot L}(\bs{u}) =\int_{\partial \Omega_N} \bs{t}_n \cdot \bs{\dot u} dA + \int_\Omega \bs{b}\cdot \bs{\dot u}d\bs{x}\,,}

\noindent where $\bs{t}_n$ are the external tractions per unit area $dA$ with outward unit normal $\bs{n}$ applied at the Neumann boundary $\partial\Omega_N$ and $\bs{b}$ are the body forces. By means of Eqs.~\ref{eq:powerF} and \ref{eq:ext_power}, Eq.~\ref{eq:eb} becomes

\eqn{eq:eb2}{\begin{split}\int_\Omega \bs{\sigma}:\bs{\dot \varepsilon}(\bs{u})+\frac{\partial \psi_{el}}{\partial d}\dot d+ f(\fpsi) \frac{G_c}{4 c_w}\left(\frac{\pfwP}{\ell}\dot d+2 \ell \nabla d \cdot \nabla \dot d\right)   d\bs{x}+\\ -\int_{\partial \Omega_N} \bs{t}_n \cdot \bs{\dot u} \,da - \int_\Omega \bs{b}\cdot \bs{\dot u}\,d\bs{x}=0\,.\end{split}}

\noindent Eq.~\ref{eq:eb2} can be integrated by parts giving the following form of the energy balance

\eqn{eq:eb3}{\begin{split}&  \int_\Omega -\left(\div\,\bs{\sigma} +\bs{b}\right) \cdot \bs{\dot u} \, d\bs{x}+  \\ & + \int_\Omega \left\{ \frac{\partial \psi_{el}}{\partial d} - \frac{G_c}{2c_w}\ell\left[f(\fpsi)\left( \Delta d- \frac{\pfwP}{2\ell^2}\right) +\nabla f(\fpsi)\cdot \nabla d\right] \right\} \dot d \, d\bs{x} +\\&+ \int_{\partial \Omega} \left( \bs{\sigma} \cdot \bs{n}- \bs{t}_n \right)\cdot\bs{\dot u} \, dA +\frac{G_c\ell}{2c_w}\int_{\partial \Omega} f(\fpsi) \, \nabla d \cdot \bs{n} \,\dot d \,  dA  = 0\,, \end{split}}

The first-order stability principle in rate form states that the solution $(\bs{u}, d)$ is stable if for any possible admissible test velocities $(\bs{ \dot{\tilde u}}, \dot{\tilde d})$  it satisfies

\eqn{eq:st1}{\begin{split}& \int_\Omega -\left(\div\,\bs{\sigma} +\bs{b}\right) \cdot \bs{\dot{\tilde u}} \, d\bs{x}+  \\ & + \int_\Omega \left[ f(\fpsi)\frac{G_c}{4c_w}\frac{\pfwP}{\ell}+\frac{\partial \psi_{el}}{\partial d} -\div\left(f(\fpsi)\frac{G_c}{2c_w}\ell \, \nabla d \right) \right] \dot{\tilde d} \, d\bs{x} + \\&  + \int_{\partial \Omega} \left( \bs{\sigma} \cdot \bs{n}- \bs{t}_n \right)\cdot\bs{\dot{\tilde u}} \, dA +\int_{\partial \Omega} f(\fpsi)\frac{G_c}{2c_w}\ell \, \nabla d \cdot \bs{n} \,\dot{\tilde d} \,  dA  \ge 0\,. \end{split}}

\noindent Note that the admissible test velocity for $d$ must satisfy the irreversibility condition Eq.~\ref{eq:irrev}.

 Using standard arguments of variational calculus, Eq.~\ref{eq:st1} leads to the following local equilibrium equation and Neumann boundary conditions for the mechanical problem

\begin{subequations}\label{eq:mech_problBC}    
\begin{equation}\label{eq:equi}
\div\,\bs{\sigma} + \bs{b}= \bs{0} \quad \text{in  } \Omega\,,
\end{equation}
\begin{equation}\label{eq:neumann}
\bs{\sigma}\cdot \bs{n} =\bs{t}_n \quad \text{on  } \partial \Omega_N\,.
\end{equation}
\end{subequations} 

\noindent (Note that $\bs{u}$ must a priori satisfy the Dirichlet boundary condition $\bs{u}=\bs{\overline u}$ on the Dirichlet boundary $\partial \Omega_D$.) For the phase-field problem the following inequalities are obtained

\begin{subequations}\label{eq:PF_probl}
\begin{equation}\label{eq:PF_evol}
 \frac{\partial \psi_{el}}{\partial d} - \frac{G_c\ell}{2c_w}\left[f(\fpsi)\left( \Delta d- \frac{\pfwP}{2\ell^2}\right) +\nabla f(\fpsi)\cdot \nabla d\right]\ge {0} \quad \text{in  } \Omega\,,
\end{equation}    
\begin{equation}\label{eq:PF_bc}
\frac{G_c\ell}{2c_w}f(\fpsi) \, \nabla d\,\cdot \bs{n} \ge0 \Rightarrow \nabla d\cdot \bs{n} \ge0 \quad \text{on  } \partial \Omega\,.
\end{equation}
\end{subequations} 

From Eq.~\ref{eq:eb3} combined with Eq.~\ref{eq:mech_problBC} the energy balance reduces to 

\eqn{eq:eb_f}{\begin{split}&  \int_\Omega \left\{ \frac{\partial \psi_{el}}{\partial d} - \frac{G_c}{2c_w}\ell\left[f(\fpsi)\left( \Delta d- \frac{\pfwP}{2\ell^2}\right)+\nabla f(\fpsi)\cdot \nabla d\right] \right\} \dot d \, d\bs{x} +\\& \qquad \qquad\qquad \qquad\qquad \qquad\qquad+ \frac{G_c\ell}{2c_w}\int_{\partial \Omega} f(\fpsi) \, \nabla d \cdot \bs{n} \,\dot d \,  dA  = 0\,. \end{split}}

\noindent Being $\dot d\ge0$ from Eq.~\ref{eq:irrev} and accounting for the inequalities in Eq.~\ref{eq:PF_probl}, the left hand side of Eq.~\ref{eq:eb_f} is the sum of two non-negative terms. Therefore, each term must vanish, which leads to the following consistency conditions

\begin{subequations}\label{eq:PF_cons}
\begin{equation}\label{eq:PF_cons1}
\left\{ \frac{\partial \psi_{el}}{\partial d} - \frac{G_c}{2c_w}\ell\left[f(\fpsi)\left( \Delta d- \frac{\pfwP}{2\ell^2}\right) +\nabla f(\fpsi)\cdot \nabla d\right] \right\}\dot d = 0 \,\, \,\text{in  } \Omega\,.
\end{equation}    
\begin{equation}\label{eq:PF_cons2}
\left( \nabla d\cdot \bs{n}\right)\dot d =0 \quad \text{on  } \partial \Omega\,.
\end{equation}
\end{subequations}

Eqs.~\ref{eq:PF_evol}, \ref{eq:irrev} and \ref{eq:PF_cons1} thus constitute the well-known Karush-Kuhn-Tucker ($KKT$) conditions which rule the evolution of the phase-field variable $d$.

	The Clausius-Duhem dissipation inequality states that the following condition on the dissipation $\mathcal{D}$ holds

\eqn{eq:diss}{\mathcal{D}=\int_\Omega W\left(\bs{\varepsilon}(\bs{u}),d,\nabla d | \fpsi\right)-\psi_{el}\left(\bs{\varepsilon}(\bs{u}), g(d)\right) d\bs{x}\ge 0\,,}

\noindent or in rate form

\eqn{eq:diss_rate}{\mathcal{\dot D}=\int_\Omega f(\fpsi) \frac{G_c}{2 c_w}\left(\frac{\pfwP}{2\ell}\dot d+\ell \, \nabla d \cdot \nabla \dot d\right)  d\bs{x}\ge0\,,}

\noindent where Eq.~\ref{eq:powerF} is accounted for. Comparing  Eq.~\ref{eq:diss_rate} with Eq.~\ref{eq:eb2} one has 

\eqn{eq:diss_rate2}{\mathcal{\dot D}=\int_\Omega -\frac{\partial \psi_{el}}{\partial d} \dot d\, d\bs{x}\ge0\,,}

\noindent which holds since $\dot d \ge 0$ (Eq.~\ref{eq:irrev}) and $\frac{\partial \psi_{el}}{\partial d}\le0$ from Eq.~\ref{eq:a_prop}.

\subsection{Cumulated history variable} \label{sct:hist_var}
As already outlined in sect.~\ref{sct:en_quantities}, $\fpsi$ is a cumulated variable which quantifies the fatigue effects already experienced by the material. 

Two cumulated variables are proposed in the following

\begin{enumerate}
\item Mean load independent: for materials whose fatigue life is not affected by the mean load of a cycle. It reads

\eqn{eq:no_mean}{\fpsi(\bs{x},t)=\int_0^t H(\alpha \dot \alpha)\,|\dot \alpha|\, d\tau\,,}

\noindent where $H(\alpha \dot \alpha)$ the Heaviside function, defined as $H(\alpha \dot \alpha)=1$ if $\alpha \dot \alpha \ge 0$ (loading) and $H(\alpha \dot \alpha)=0$ otherwise (unloading).

\item Mean load dependent: the model can be enriched by introducing a history variable that weighs differently the rate of the cumulated variable depending on the load level achieved as

\eqn{eq:mean}{\fpsi(\bs{x},t)=\frac{1}{\alpha_N}\int_0^t  H(\alpha\dot \alpha) \, \alpha \,\dot \alpha\, d\tau\,,}

\noindent where $\alpha_N$ is a normalization parameter needed to achieve dimensional consistency.
\end{enumerate}

\subsection{Fatigue degradation function} \label{sct:hist_var}
The fatigue degradation function $f(\fpsi(t))$ describes how fatigue effectively reduces the fracture toughness of the material. The following two fatigue degradation functions are considered here \citep{Alessi2017}

\begin{figure}[!h]
	\begin{adjustwidth}{-3cm}{-3cm}
	\centering
		\includegraphics{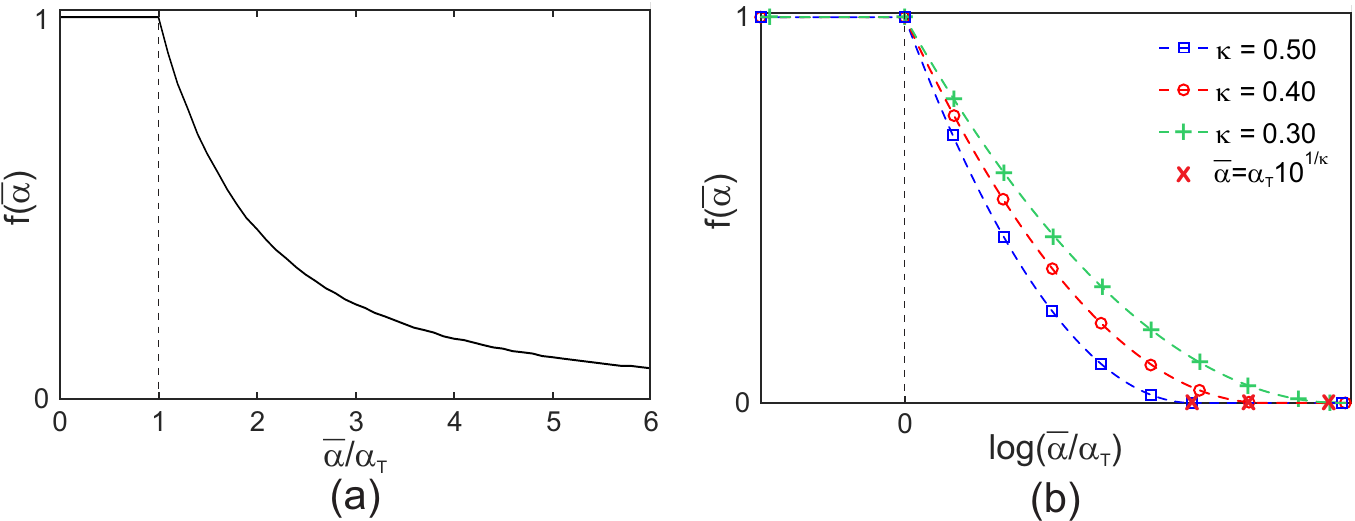}                                                                                             
	\end{adjustwidth}
		\caption{Fatigue degradation function: (a) asymptotic and (b) logarithmic for different values of $\kappa$.} 
		\label{fig:degr_fct}
	\end{figure}

\eqn{eq:f_basic}{f(\fpsi(t))=
\begin{cases}
1 & \text{if}\quad \fpsi(t) \le\alpha_T \\
\displaystyle \left( \frac{2\alpha_T}{\fpsi(t)+\alpha_T}\right)^2 &  \text{if}\quad \fpsi(t) \ge \alpha_T \,,
\end{cases}}

\noindent and
\eqn{eq:f_log}{f(\fpsi(t))=
\begin{cases}
1 & \text{if}\quad \fpsi(t) \le\alpha_T \\
\displaystyle \left[1- \kappa \log\left(\frac{\fpsi(t)}{\alpha_T}\right)\right]^2 &  \text{if}\quad \alpha_T\le\fpsi(t) \le \alpha_T10^{1/\kappa} \\
0 &  \text{if}\quad \fpsi(t) \ge\alpha_T10^{1/\kappa}
\end{cases}}

\noindent where $\kappa$ is a material parameter. The difference between Eqs.~\ref{eq:f_basic} and \ref{eq:f_log} is that the former (Fig.~\ref{fig:degr_fct}a) delivers an asymptotically vanishing value while the latter (Fig.~\ref{fig:degr_fct}b) vanishes for a finite value of $\fpsi(t)$. Also, the slope of the logarithmic function Eq.~\ref{eq:f_log} can be tuned by varying $\kappa$ as shown in Fig.~\ref{fig:degr_fct}b, giving a further degree of freedom when simulating real material behaviors.

\subsection{Choice of $\alpha$} \label{sct:alpha_choice}
Concerning the fatigue history variable, in \citep{Alessi2017} it is assumed that the fatigue effects are driven by the strain, namely $\alpha=\varepsilon$ (in 1-D), which could be extended to 2- or 3-D by taking $\alpha = ||{\boldsymbol{\varepsilon}}||$. Although proven to be effective in 1-D, this solution gives rise to mesh-dependency issues in a multi-dimensional framework as illustrated in sect.~\ref{sct:num_fatigue}. Due to the energetic nature of the adopted modeling framework, it seems natural to account for the active part of the elastic strain energy density, i.e. $\alpha = \psi^+_{el}(\bs{\varepsilon},d)$. Also, the fatigue effects are cumulated only during the loading phase \citep{Alessi2017,Jaubert2006}, defined as  $ \dot \psi_{el}^+\ge0$.

\subsection{Summary} \label{sct:summary}
The governing equations and needed parameters of the proposed modeling framework are summarized in Tab.~\ref{tab:equations} where $\mathbbm{C}^+(\lambda,\nu)$ and $\mathbbm{C}^-(\lambda,\nu)$ are the constitutive elastic tensors related to the active and inactive parts of the elastic strain energy density.

		\begin{table}\footnotesize
	\begin{adjustwidth}{-3cm}{-3cm}
		\centering
	\begin{tabular}{P{3cm}P{12.2cm}} \toprule
		\multicolumn{2}{c}{\textbf{Balance of momentum}}\\ \toprule
		\vspace{-5mm}\textbf{Equilibrium equation} &\vspace{-5mm}\eqnn{\div\left\{ \left[g(d)\mathbbm{C}^+(\lambda,\nu)+\mathbbm{C}^-(\lambda,\nu)\right]:\bs{\varepsilon}\right\}+ \bs{b}= \bs{0} \quad \text{in  } \Omega\,,} \\[-3em]
		\textbf{Boundary conditions} &\begin{subequations}
\begin{equation*}\label{eq:neumann}
\bs{\sigma}\cdot \bs{n} =\bs{t}_n \quad \text{on  } \,\,\partial \Omega_N\,,
\end{equation*}
\begin{equation*}\label{eq:dirichlet}
\bs{u} = \bs{\overline u} \quad \text{on  }\,\, \partial\Omega_D\,,
\end{equation*}
\end{subequations}  \\[-3em]
\textbf{Degradation function} & \eqnn{g(d)=(1-d)^2}  \\[-1em]
\textbf{Parameters} & {$\lambda$, $\nu$ $\rightarrow$ $\mathbbm{C}^+(\lambda,\nu)$, $\mathbbm{C}^-(\lambda,\nu)$}\\ \toprule

\multicolumn{2}{c}{\textbf{Phase-field evolution}}\\ \toprule
		\vspace{-5mm}\textbf{KKT conditions} & \vspace{-5mm}\begin{equation*}\begin{cases}
 \frac{\partial \psi_{el}}{\partial d}-\frac{G_c}{2c_w}\ell\left[f(\fpsi)\left( \Delta d- \frac{\pfwP}{2\ell^2}\right) +\nabla f(\fpsi)\cdot \nabla d\right]\ge 0\,,\\
 \dot d \ge 0\,,  \\
\left \{ - \frac{\partial \psi_{el}}{\partial d} + \frac{G_c}{2c_w}\ell\left[f(\fpsi)\left( \Delta d- \frac{\pfwP}{2\ell^2}\right) +\nabla f(\fpsi)\cdot \nabla d\right]\right\}\dot d = 0\,.\\

\end{cases} 
\end{equation*} \\[-3em]
		\textbf{Boundary conditions} & \eqnn{\nabla d \cdot \bs{n} = 0 \quad \text{on  } \partial \Omega}\\[-1em]
				\textbf{Dissipation function} & $\pfw=d^2$ (\texttt{AT2}) \;\;\;or\;\;\; $\pfw=d$ (\texttt{AT1})  \\
\vspace{2mm}\textbf{Parameters} &\vspace{2mm} {$G_c$, $\ell$ }\\ \toprule
	
	\multicolumn{2}{c}{\textbf{Fatigue}}\\ \toprule
		\multirow{2}{*}{\textbf{History variable}} &  $\dot{\overline\alpha} =H(\dot{\psi_ {el}^+})\dot{\psi_{el}^+}$ \;\;\; (No mean load eff.)\\[1em]
												& $\displaystyle\dot{\overline\alpha} =\frac{1}{\alpha_N}H(\dot{\psi_{el}^+})\psi_{el}^+\dot{\psi_{el}^+}$ \;\;\; (Mean load eff.)$^{\dagger}$ \\[1.5em]
		{\textbf{Fatigue degradation function}} & $f(\fpsi(t))=
\begin{cases}
1 & \text{if}\quad \fpsi(t) \le\alpha_T \\
\displaystyle \left( \frac{2\alpha_T}{\fpsi(t)+\alpha_T}\right)^2 &  \text{if}\quad \fpsi(t)\ge\alpha_T \,,
\end{cases} \quad\textrm{(Asymptotic)}$

\vspace{4mm}$f(\fpsi(t))=
\begin{cases}
1 & \text{if}\quad \fpsi(t) \le\alpha_T \\
\displaystyle \left[1- \kappa \log\left(\frac{\fpsi(t)}{\alpha_T}\right)\right]^2 &  \text{if}\quad \alpha_T\le\fpsi(t) \le \alpha_T10^{1/\kappa} \\
0 &  \text{if}\quad \fpsi(t) \ge \alpha_T10^{1/\kappa}
\end{cases}\quad \textrm{(Logarithmic})^{\square}$ \\ [3em]
\textbf{Parameters} & {$\alpha_T$, $\alpha_N$ (only for $\dagger$), $\kappa$ (only for $\square$) }\\ \toprule
	\end{tabular}
	\end{adjustwidth}
		\caption{Governing equations and parameters. } 
		 \label{tab:equations}
	\end{table}


\section{Numerical implementation} \label{sct:comp_asp}
To find the numerical solution, the mechanical and phase-field  problems are written in weak form and, once discretized using linear finite elements and recasted in incremental form, they are solved using a staggered approach \citep{Miehe2010}. The convergence of the solution is ensured controlling the residual as proposed in \citep{Ambati2014}. As in \citep{Miehe2010}, the crack irreversibility condition is enforced introducing the history variable

\eqn{eq:hist_irr}{\mathcal{H}=\max_{\tau\in[0,t]}\psi_{el,0}^+(\bs{\varepsilon}({\bf x},\tau)) \,.}

\noindent The alternative approaches are discussed in \citep{Timofy}.

\subsection{Time integration}

Within the time discretized setting, the updated value of the fatigue history variable reads

\eqn{142}{\overline\alpha_{n + 1} =  {\overline\alpha_n} +\int_{t_n}^{t_{n+1}}\dot{\overline\alpha} \, d\tau = {\overline\alpha_n} + \Delta{\overline\alpha} \, ,}

\noindent where the subscripts $_{n}$ and $_{n+1}$ refer to the instants $t=t_n$ and $t=t_{n+1}$, respectively, $\Delta t=t_{n+1}-t_n$  and $\Delta{\overline\alpha}$ is approximated as
 
\eqn{143}{\Delta{\overline\alpha}  = \left|\alpha_{n+1}- {\alpha_{n}}\right| H\left(\frac{\alpha_{n+1}- {\alpha_{n}}}{\Delta t}\right)\, ,}

\noindent or  
\eqn{144}{\Delta{\overline\alpha}  = \frac{ \left(\alpha_{n+1}- {\alpha_{n}}\right)}{\alpha_N} \left(\frac{\alpha_{n +1}+ {\alpha_{n}}}{2}\right) H\left(\frac{\alpha_{n+1}- {\alpha_{n}}}{\Delta t}\right)\, ,}

\noindent respectively if the mean load independent (Eq.~\ref{eq:no_mean}) or dependent (Eq.~\ref{eq:mean}) accumulation function is used.


\section{Numerical examples} \label{sct:num_ex}

In this section some numerical experiments are presented and discussed. On one hand we illustrate the major features of the proposed framework and on the other hand we compare qualitatively and quantitatively the numerical results with the known results about fatigue.

In the remainder of this paper and if not specified otherwise, the \texttt{AT2} model (Eq.~\ref{eq:w_AT2}), the mean load independent accumulation function (Eq.~\ref{eq:no_mean}) and the asymptotic fatigue degradation function (Eq.~\ref{eq:f_basic}) are used, while a spatial discretization $h = \frac{\ell}{5}$ is adopted within the region of the specimen where the crack propagation is expected. Furthermore, when calculating the crack length,  the origin is taken at the pre-existing notch tip and we consider completely cracked any point where $d\ge$ 0.95. 

The fatigue-related quantities $\alpha_T$ and $\alpha_N$ are meant to be material parameters to be determined on an experimental basis. However, the comparison with experimental results is out of the scope of the present work. Hence, it is here assumed that 

\eqn{eq:alpha_T}{\alpha_T=\alpha_N=\frac{1}{2}\varepsilon_y^{AT2}E\varepsilon_y^{AT2}\,,}  

\noindent where $\varepsilon_y^{AT2}$ is defined in Eq.~\ref{eq:hom_AT2}. This choice is made to permit a proper comparison between the results as well as to highlight that no fine tuning of the parameter is performed.

	\subsection{Single-edge notched test} \label{sct:num_fatigue}
\begin{figure}[!h]
	\begin{adjustwidth}{-3cm}{-3cm}
	\centering
		\includegraphics{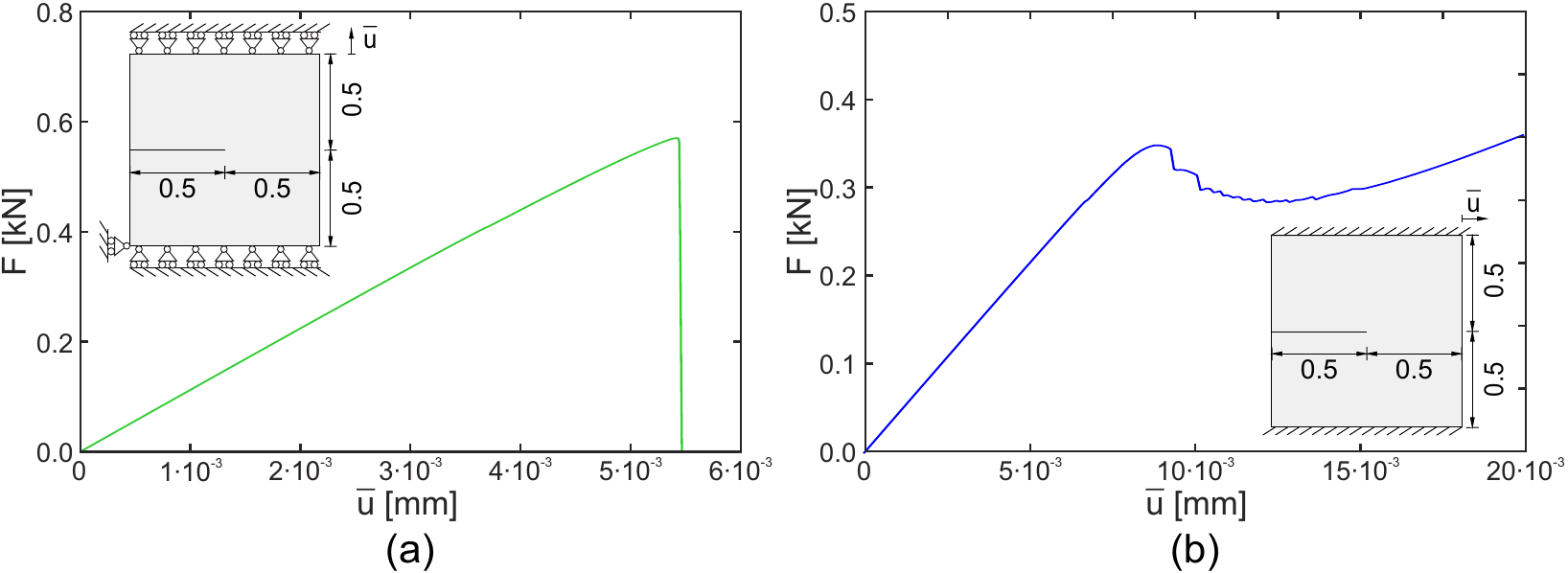}                                                                                             
	\end{adjustwidth}
		\caption{Geometry and boundary conditions of the single edge notched specimen: (a) tensile and (b) shear test.} 
		\label{fig:sent_geom}
	\end{figure}

	To evaluate the performance of the model, the widely investigated single-edge notched specimen \citep{Ambati2014,Miehe2010,Steinke2018} is tested. Pure tensile and shear conditions are considered and the corresponding numerical monotonic loading curves along with the geometry and boundary conditions used are presented in Fig.~\ref{fig:sent_geom}. The material parameters adopted are $E=$ 210GPa, $\nu=0.3$, $G_c=2.70$N/mm, $\alpha_T=\alpha_N=5.625\cdot10^{1}$N/mm$^2$ and $\ell=0.004$mm. All the simulations are performed in plane strain conditions and displacement control. 
	
	Note that the maximum monotonic load obtained using the present framework is slightly lower than that obtained with the standard formulation of sect.~\ref{sct:monotonic} if the same elastic and phase-field parameters are used. This happens because already during the monotonic loading phase a certain effective degradation of $G_c$ takes place. This only apparent issue can be solved by a proper parameter calibration. Also, some jumps in the load-displacement curve due to locally induced snap-back instabilities are observed in the the post-peak phase of the shear test (Fig.~\ref{fig:sent_geom}b). The same issue is observed also with the standard monotonic approach, although to a different extent. This difference can be ascribed to the effect of $f(\fpsi)$ that degrades the fracture toughness of the portion of material ahead of the crack tip, which leads to phases of abrupt crack propagation.

\begin{figure}[!h]
	\begin{adjustwidth}{-3cm}{-3cm}
	\centering
		\includegraphics{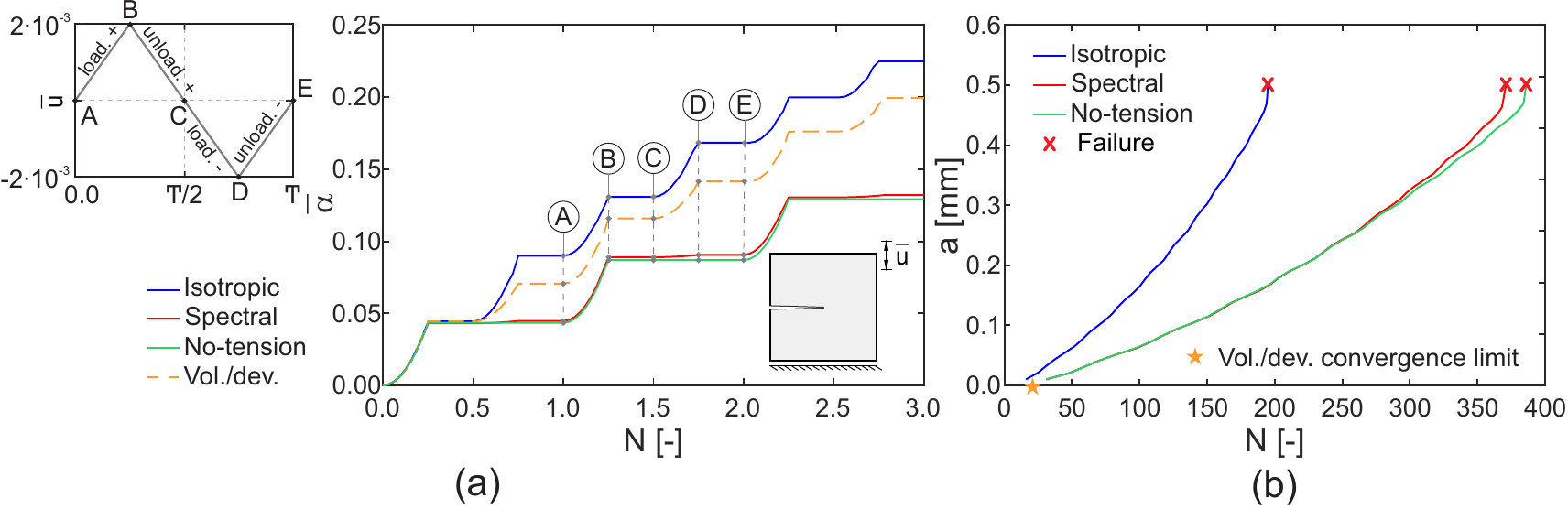}                                                                                             
	\end{adjustwidth}
		\caption{Accumulation of the fatigue history variable $\alpha$ for few illustrative cycles and (b) fatigue life curves up to failure for different splits.} 
		\label{fig:sent_split}
	\end{figure}

	To investigate the effect of the adopted tension-compression split (sect.~\ref{sct:el_en}), a cyclic tensile test is performed. A symmetric cyclic load is applied with $\Delta \overline u$ = 4$\cdot$10$^{-3}$mm (Fig.~\ref{fig:sent_split}a). The results in terms of the accumulation of the fatigue history variable $\overline \alpha$ vs. the number of cycles $N$ (fatigue life curves) illustrate that the isotropic model \citep{Bourdin2000} leads to the evolution of $\alpha$ during all the loading phases, regardless of weather the induced stress state is positive, i.e. tension (branch \textsf{AB} in Fig.~\ref{fig:sent_split}a), or negative, i.e. compression (branch \textsf{CD} in Fig.~\ref{fig:sent_split}a). The consequence is an unphysical behavior for both fracture and fatigue, since a crack is meant to propagate under tension \citep{Amor2009,Miehe2010,Freddi2010} and it was proven by Elber \citep{Elber1970} that, in case of negligible plasticity, the detrimental fatigue effect should be attributed to tensile stress states only. 

	Adopting the volumetric/deviatoric split \citep{Amor2009} leads to loss of iterative convergence after the onset of the crack (Fig.~\ref{fig:sent_split}b). This is due to a pathological behavior of the split itself that leaves undamaged only the compressive volumetric strain energy and thus leads to a fluid-like behavior. For this reason the volumetric/deviatoric split is no longer considered in the following. The spectral \citep{Miehe2010} and no-tension \citep{Freddi2010} splits distinguish between tension and compression loading phases and behave similarly, the only difference being the amount of energy dissipated during the negative loading phases and hence the evolution of $\overline \alpha$. In particular for the case analyzed here, the spectral split degrades a limited fraction of the deviatoric energy related to the positive principal strain, while the no-tension split does not degrade energy. This leads to negligible differences in the fatigue life, whereas using the isotropic model the fatigue life is decreased by half being the degradation double (Fig.~\ref{fig:sent_split}b). Note that, for all models, the accumulation of the fatigue history variable $\overline \alpha$ during unloading (branches \textsf{BC} and \textsf{DE} in Fig.~\ref{fig:sent_split}a) is prevented by the loading-unloading condition of Eqs.~\ref{eq:no_mean} and \ref{eq:mean}. 

	Following the obtained results, in the remaining of the paper the spectral split will be used if not differently specified.	
			
\begin{figure}[!h]
	\begin{adjustwidth}{-3cm}{-3cm}
	\centering
		\includegraphics{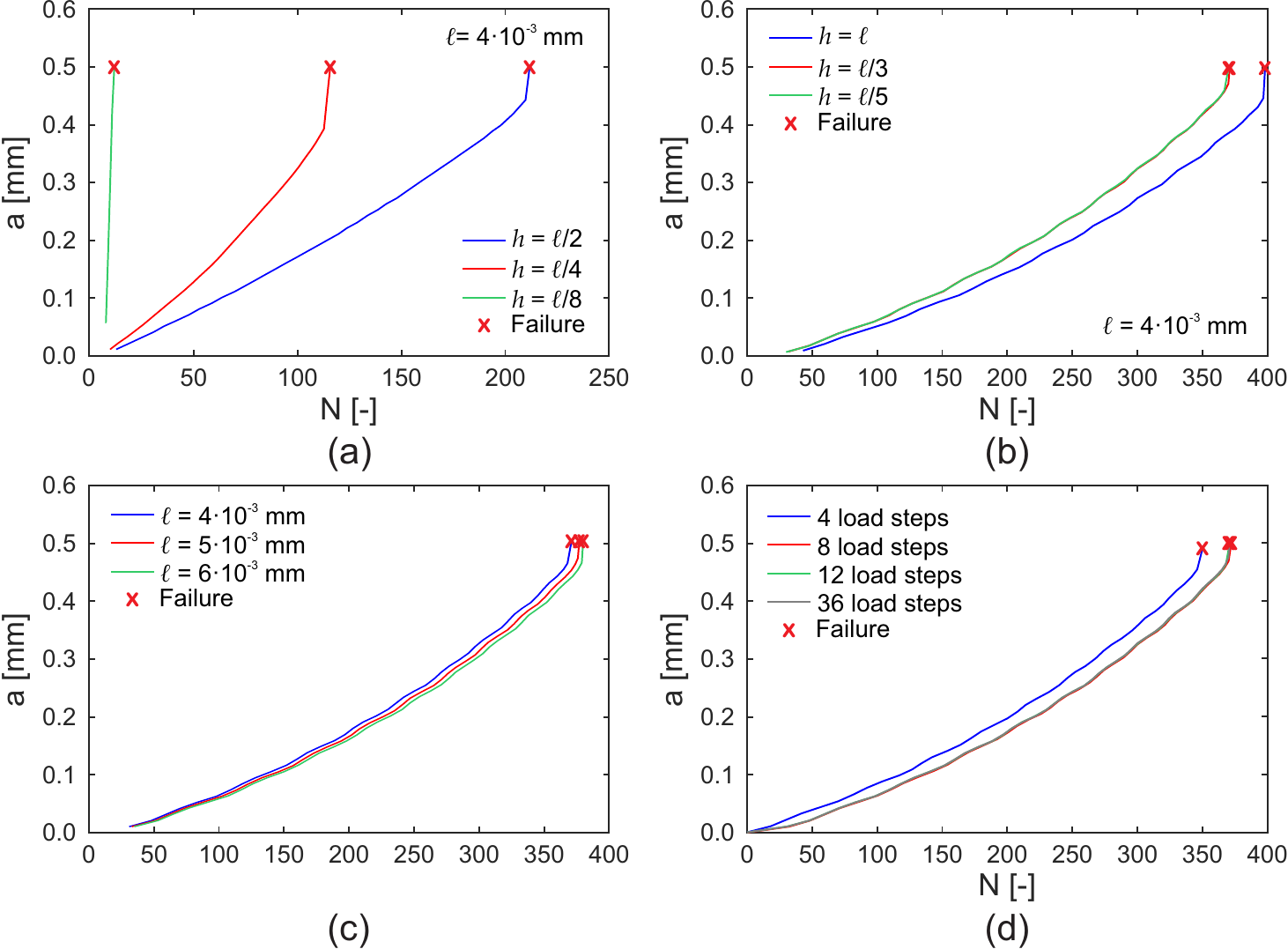}                                                                                             
	\end{adjustwidth}
		\caption{Crack length $a$ vs. number of cycles $N$: convergence study for different mesh sizes $h$ adopting (a) $\alpha=|| \bs{\varepsilon}||$ or (b) $\alpha=\psi_{el}^+$,  (c) effect of the length scale parameter $\ell$ and (d) effect of the cycle (i.e., pseudo-time) discretization.} 
		\label{fig:sent_conv}
	\end{figure}

		The convergence of the simulations with the mesh size $h$ using either $\alpha=|| \bs{\varepsilon}||$ or $\alpha=\psi_{el}^+$ is evaluated in Figs.~\ref{fig:sent_conv}a,b. As mentioned earlier, the former choice leads to a strong mesh dependence (Fig.~\ref{fig:sent_conv}a) due to the strain field singularity at the crack tip. Conversely, with $\alpha=\psi_{el}^+$ convergence is already reached for $h=\ell/3$ (Fig.~\ref{fig:sent_conv}b). The sensitivity of the results to the length scale parameter $\ell$ is investigated in Fig.~\ref{fig:sent_conv}c. Here we can observe that, in agreement with the results in \citep{Tanne2018}, only marginal differences are visible varying the length scale from 6$\cdot$10$^{-3}$mm to 4$\cdot$10$^{-3}$mm. Fig.~\ref{fig:sent_conv}d illustrates the results of the simulations for different pseudo-time discretizations of the load cycle \textsf{ABCDE} of Fig.~\ref{fig:sent_split}a, demonstrating that convergence is reached already using 8 load steps per cycle. 
		
\begin{figure}[!h]
	\begin{adjustwidth}{-3cm}{-3cm}
	\centering
		\includegraphics{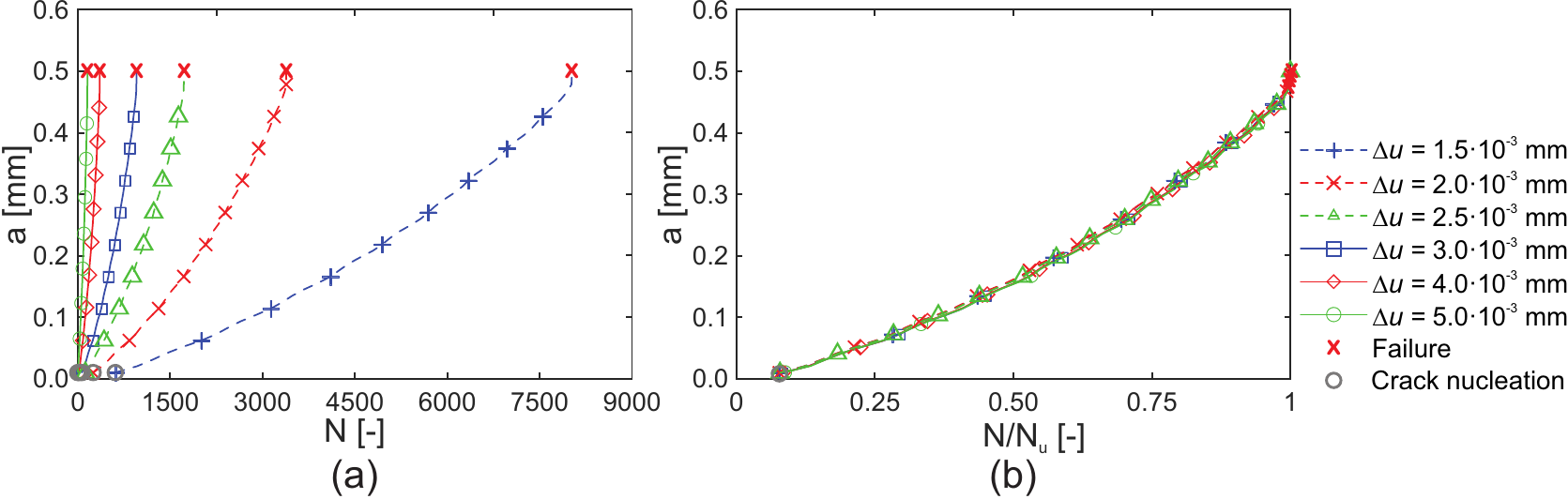}                                                                                             
	\end{adjustwidth}
		\caption{Fatigue life curves for different applied displacement ranges $\Delta u$ (a) $N$ vs. $a$ and (b) normalized $N/N_u$ vs. $a$.} 
		\label{fig:miner}
	\end{figure}

 Fig.~\ref{fig:miner}a shows the substantial role of the load range on the fatigue life as predicted by the proposed model. Increasing the load range leads to a reduction of number of cycles to both crack nucleation $N_y$ and failure $N_u$. Interestingly, upon normalizing the result by $N_u$ all curves cluster together, meaning that the fatigue process depends mainly on how far the crack has propagated rather than on the load history. This is due to the specific displacement controlled boundary condition of Fig.~\ref{fig:sent_geom}a, that makes the product between the applied force and the shape factor of the stress intensity factor range roughly constant (see \ref{app:miner_appr} for a detailed justification). 

\begin{figure}[!h]
	\begin{adjustwidth}{-3cm}{-3cm}
	\centering
		\includegraphics{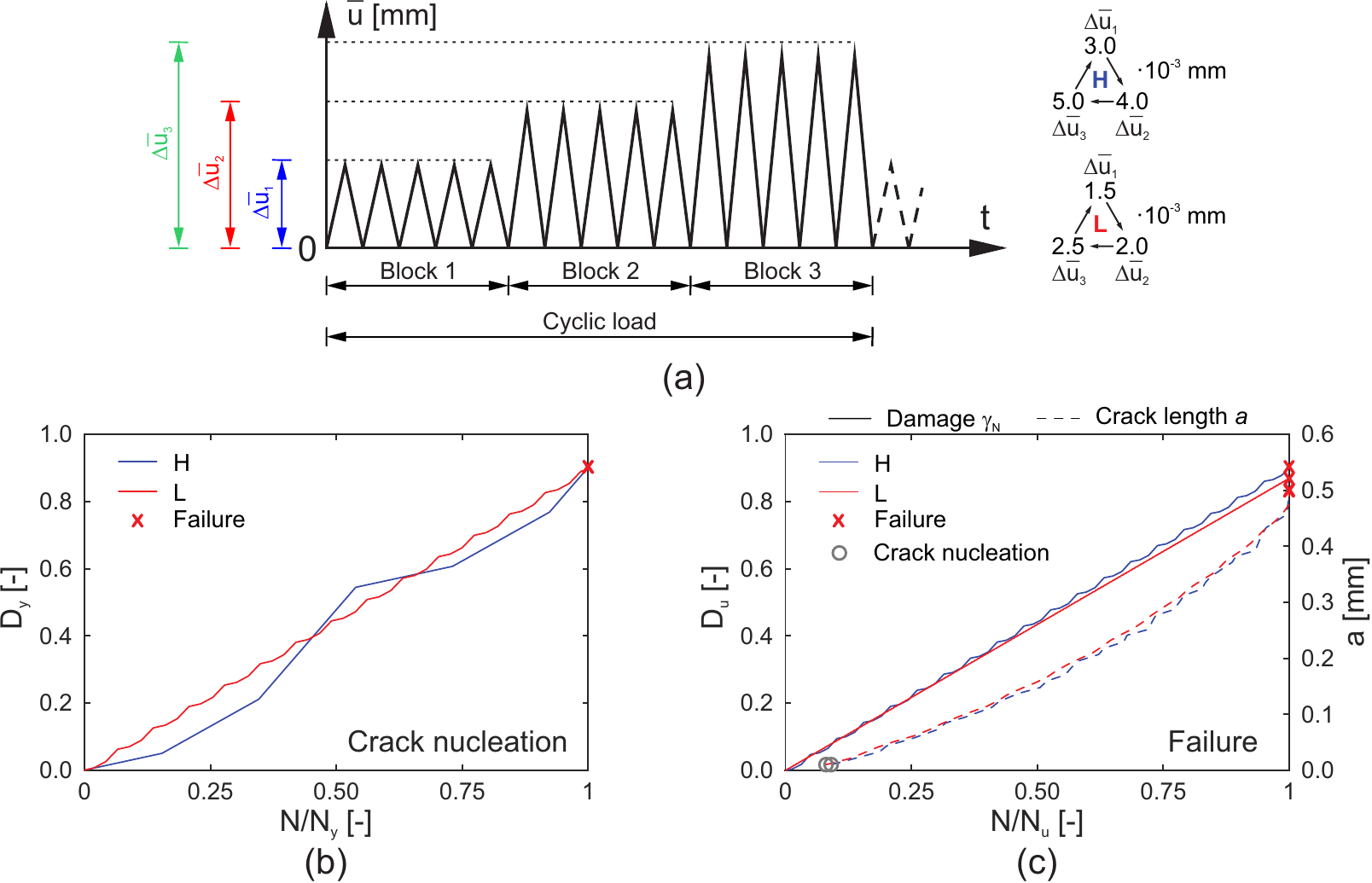}                                                                                             
	\end{adjustwidth}
		\caption{(a) Sequential multiblock cyclic load and accumulation of the Palmgren-Miner-type damage $\gamma_N$ for (b) crack nucleation and (c) failure (along with the final fatigue life curve).} 
		\label{fig:multibl}
	\end{figure}

The aforementioned condition is in agreement with the assumptions underlying Eq.~\ref{eq:miner}, suggesting thus that the Palmgren-Miner criterion can be approximately reproduced by the present approach for the special case of the displacement controlled tests. To prove this, two tests are performed where the cyclic load is constituted by three blocks of five constant amplitude cycles as in Fig.~\ref{fig:multibl}a. The ranges of the applied displacement are $\Delta \overline u_1$ = 3.0$\cdot$10$^{-3}$mm - $\Delta \overline u_2$ = 4.0$\cdot$10$^{-3}$mm - $\Delta \overline u_3$ = 5.0$\cdot$10$^{-3}$mm for the test labeled \textsf{H} and  $\Delta \overline u_1$ = 1.5$\cdot$10$^{-3}$mm - $\Delta \overline u_2$ = 2.0$\cdot$10$^{-3}$mm - $\Delta \overline u_3$ = 2.5$\cdot$10$^{-3}$mm for the test labeled \textsf{L}. Figs.~\ref{fig:multibl}b,c present the cumulated Palmgren-Miner damage $D=\sum_{i=1}^N\gamma_i$ respectively for crack nucleation $D_y$ and failure $D_u$ of the specimen. Both approach the unit value with deviations around 10\%. As highlighted in \ref{app:miner_appr}, this is one of the few special cases where the assumptions of the Palmgren-Miner rule are approximately satisfied for a crack growth ruled by Paris law.

\begin{figure}[!h]
	\begin{adjustwidth}{-3cm}{-3cm}
	\centering
		\includegraphics{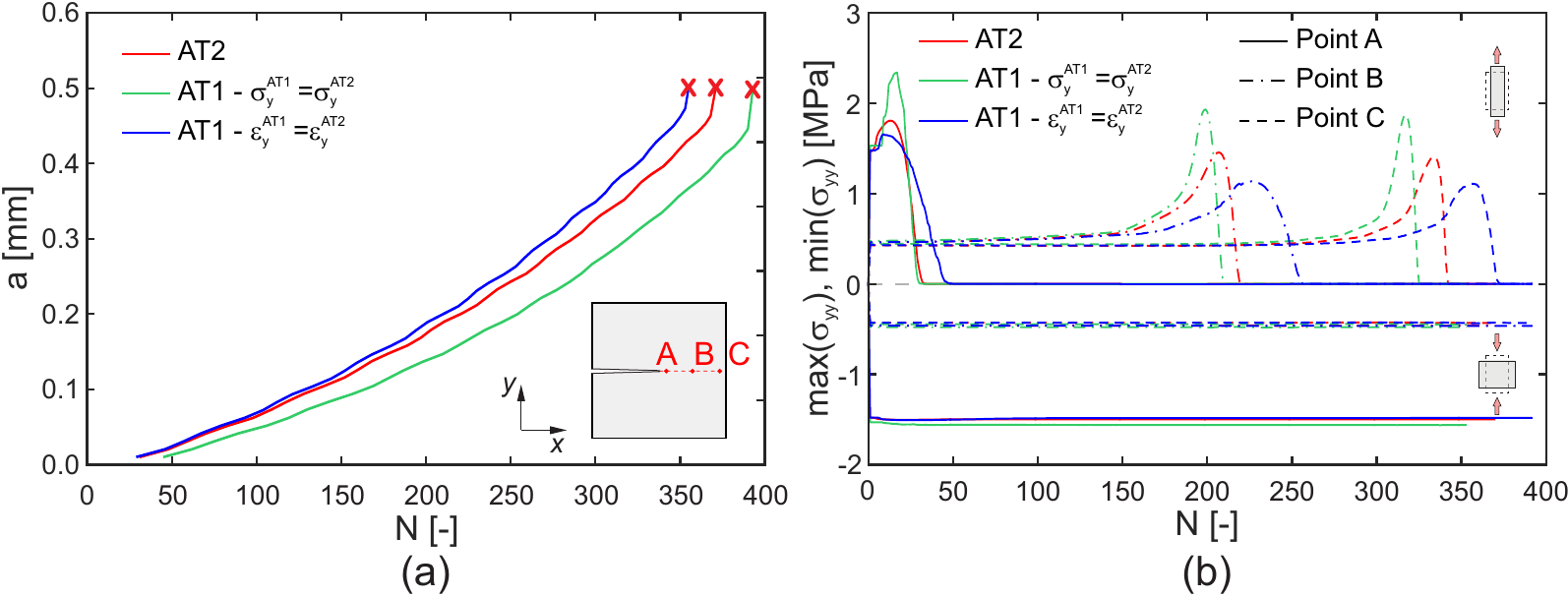}                                                                                             
	\end{adjustwidth}
		\caption{(a) Fatigue life curves adopting the \texttt{AT2} or \texttt{AT1} model and (b) maximum and minimum value of $\sigma_{yy}$ vs. cycle number $N$.} 
		\label{fig:sent_maxs}
	\end{figure}

		 Fig.~\ref{fig:sent_maxs}a compares the fatigue life curves obtained adopting the \texttt{AT2} and \texttt{AT1} models. For a proper comparison between the two models, $E$ and $G_c$ are kept constant while the regularization length $\ell$ is determined either by imposing $\sigma_y^{AT2}=\sigma_y^{AT1}$ or $\varepsilon_y^{AT2}=\varepsilon_y^{AT1}$ following Eqs.~\ref{eq:hom_AT2} and \ref{eq:hom_AT1}. The difference in terms of fatigue life $N_u$ is limited and ranges between 10$\%$ and 15$\%$.
		
	 More differences can be observed by comparing the local behavior. Fig.~\ref{fig:sent_maxs}b shows the maximum and minumum value of $\sigma_{yy}$ reached in each cycle  for the points \textsf{A}, \textsf{B} and \textsf{C} sketched in Fig.~\ref{fig:sent_maxs}a and located at the same height of the notch and at 0.01, 0.20 and 0.40 mm distance from its tip. Close to the notch tip, the accumulation of the fatigue history variable has less influence on the strength of the material because the softening phase is reached during the early stages of the test (Fig.~\ref{fig:sent_maxs}b). Conversely, away from the notch tip, $\overline\alpha$ can increase significantly before the softening phase, leading to lower maximum stresses attainable.

\begin{figure}[!h]
	\begin{adjustwidth}{-3cm}{-3cm}
	\centering
		\includegraphics{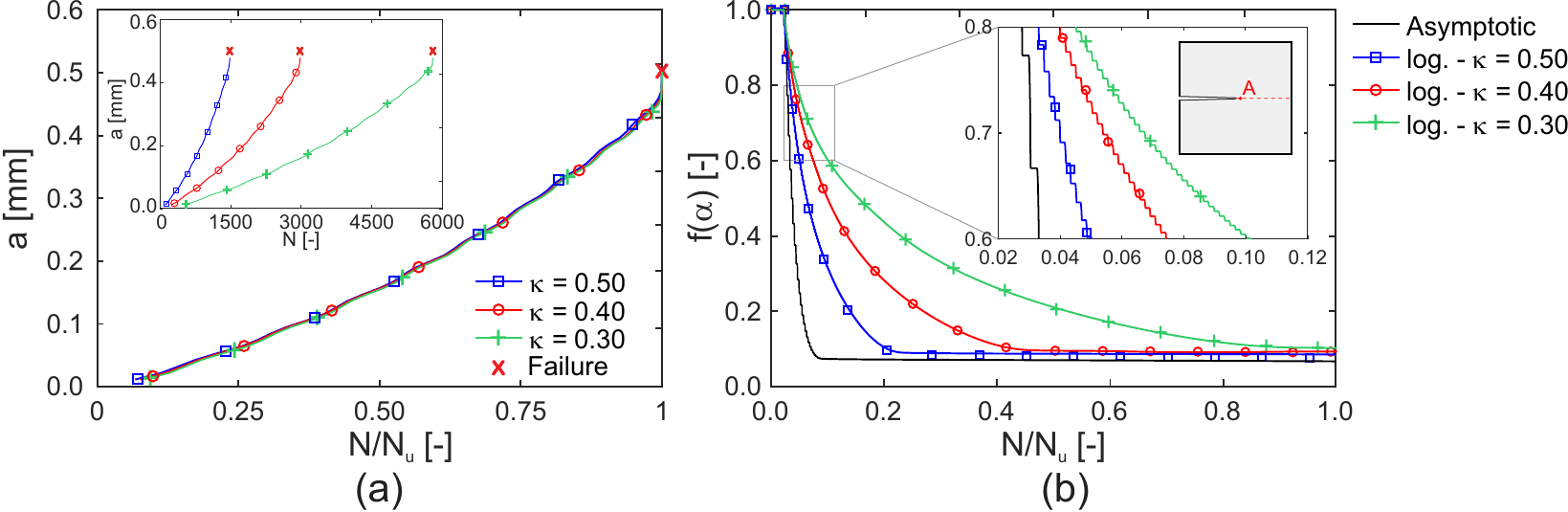}                                                                                             
	\end{adjustwidth}
		\caption{Fatigue life prediction using the logarithmic fatigue degradation function $f(\alpha)$ (Eq.~\ref{eq:f_log}): (a) fatigue life curves for different $\kappa$ parameters and (b) pertaining evolution of $f(\alpha)$ and comparison with the asymptotic fatigue degradation function (Eq.~\ref{eq:f_basic}).} 
		\label{fig:sent_logfct}
	\end{figure}

		Fig.~\ref{fig:sent_logfct}a shows the results obtained adopting the logarithmic fatigue degradation function Eq.~\ref{eq:f_log} for the same values of the material parameter $\kappa$ of Fig.~\ref{fig:degr_fct}b. As expected, increasing the slope of the fatigue degradation function leads to a shorter fatigue life. However, if the number of cycles is normalized by $N_u$, we obtain again a single fatigue life curve. The evolution of the asymptotic and logarithmic degradation functions for the point \textsf{A} of Fig.~\ref{fig:sent_maxs}a are compared in Fig.~\ref{fig:sent_logfct}b. Here we can observe an initial phase where the functions are constant and equal to 1, followed by a decreasing branch that ends with a plateau, starting when the phase field variable in \textsf{A} attains the unity and the fatigue history variable cannot evolve anymore. 
		
\begin{figure}[!h]
	\begin{adjustwidth}{-3cm}{-3cm}
	\centering
		\includegraphics{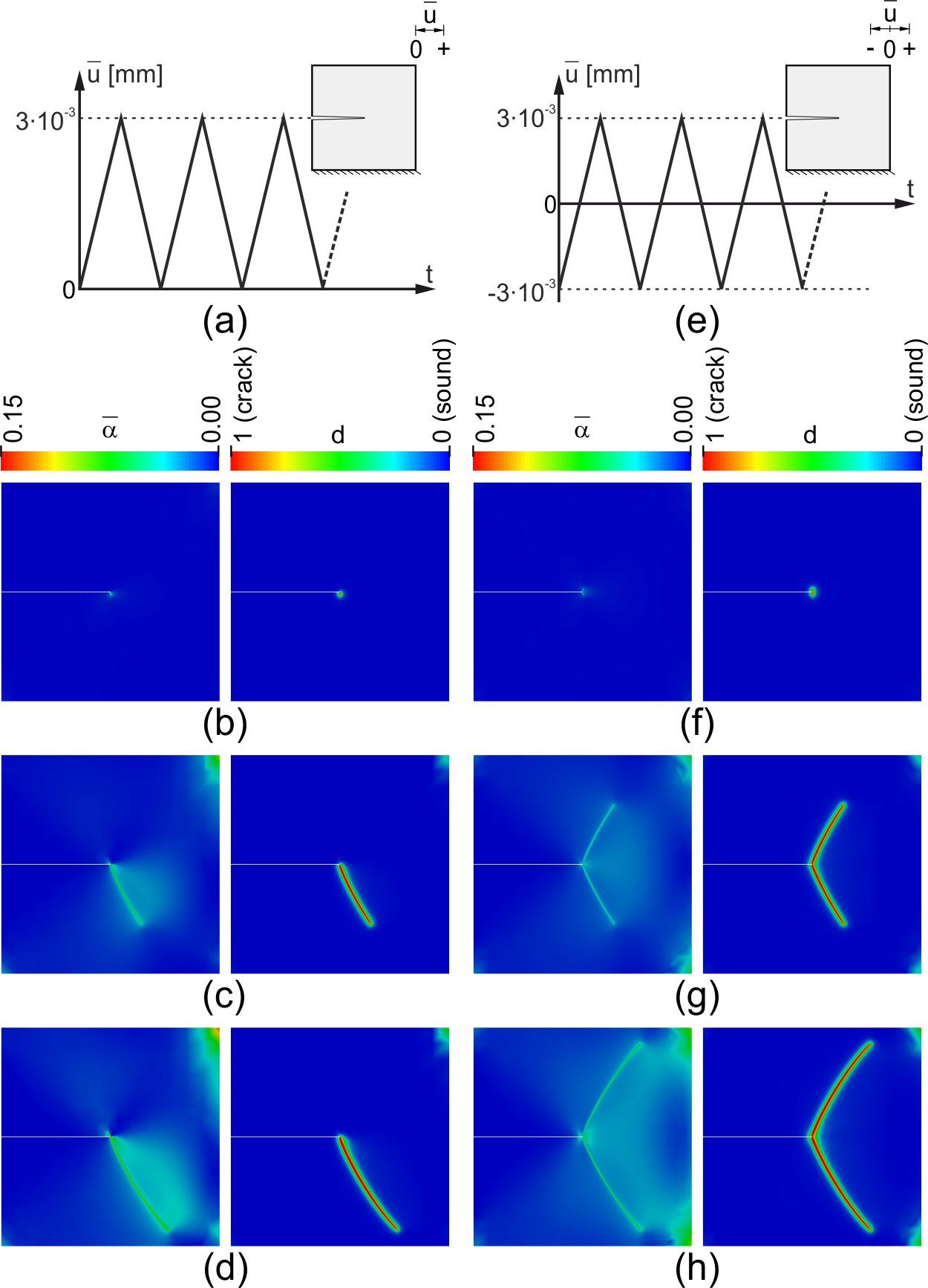}                                                                                             
	\end{adjustwidth}
		\caption{Fatigue shear test for the single-edge notched specimen. (a) Non-inverting applied displacement and (b), (c), (d) fatigue history variable and phase-field maps respectively for crack nucleation, stable propagation and incipient failure. (e) Symmetrically inverting applied displacement and (f), (g), (h) fatigue history variable and phase-field maps respectively for crack nucleation, stable propagation and incipient failure.} 
		\label{fig:sent_shear}
	\end{figure}

	A further test is performed by subjecting the single-edge notched specimen to shear loading, see Fig.~\ref{fig:sent_geom}b. First, a non-inverting displacement is applied with range $\Delta \overline u$ = 3$\cdot$10$^{-3}$mm (Fig.~\ref{fig:sent_shear}a). The results in terms of cumulated history and phase-field variables are reported in Figs.~\ref{fig:sent_shear}b,c~and~d respectively for nucleation, stable and unstable propagation of the crack.  Here we can observe that, although the fatigue history variable $\overline \alpha$ is accumulated in the whole domain over large regions, the observed crack maintains its localized nature and its path is similar to the one observed in the monotonic case, namely it originates at the initial notch tip and propagates downwards toward the lower right corner (Figs.~\ref{fig:sent_shear}b-d). A second test is performed by applying to the same specimen a symmetrically inverting displacement with range  $\Delta \overline u$ = 6$\cdot$10$^{-3}$mm as shown in Fig.~\ref{fig:sent_shear}e. In this case the accumulation of $\overline \alpha$ and the evolution of the phase field variable $d$ are different from the previous example. Two cracks branch from the tip of the initial notch and propagate symmetrically toward the top and bottom corners at the right-hand side (Figs.~\ref{fig:sent_shear}f-h). In particular, the upper and lower cracks are the result respectively of the negative and positive part of the cyclic load (Fig.~\ref{fig:sent_shear}e). The accumulation of $\overline \alpha$ spreads over even larger regions (Figs.~\ref{fig:sent_shear}f-h) but, again, the crack remains localized. This example also demonstrates the ability of the approach to handle branching cracks. 
		
	\subsection{Comparison with Paris theory} \label{sct:Paris}
	In this section we compare the results of the proposed approach with the main features of the Paris theory in terms of both crack growth rate curves and parameters of the Paris law $C$ and $m$. In particular, the importance of the Paris law lies in the fact that $C$ and $m$ are meant to be material parameters\footnote{In case of non-linear fatigue behaviors the Paris parameters are no longer constant but, rather, functions of certain quantities. E.g., for mean load sensitive materials they are related to the mean load, see sect.~\ref{sct:CT_ex}.}. Such condition is not met by the majority of the available approaches to fatigue, such as the W{\"o}hler curve for which the parameters of the Basquin relationship (Eq.~\ref{eq:basquin}) change with varying geometry and boundary conditions. To test whether our approach recovers the Paris theory and whether the corresponding parameters depend or not on the test setup,  we analyze and compare the results obtained simulating two popular tests to characterize the fatigue behavior of a brittle material, i.e. the \emph{compact tension} (\emph{CT}) and the \emph{three-point-bending} (\emph{TPB}) tests.

\begin{figure}[!h]
	\begin{adjustwidth}{-3cm}{-3cm}
	\centering
		\includegraphics{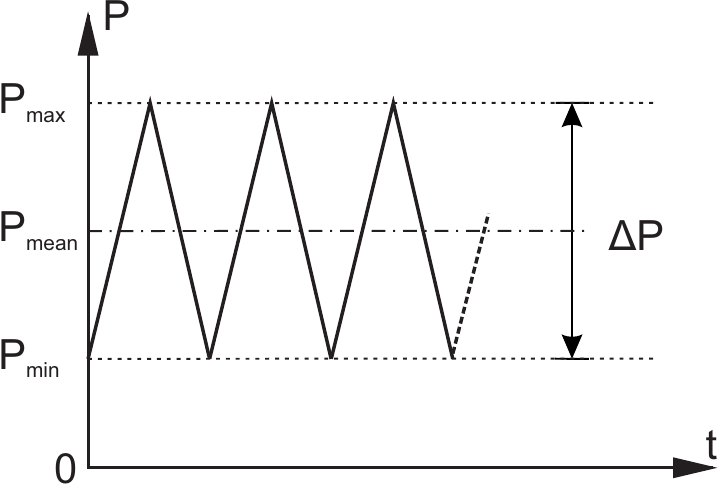}                                                                                             
	\end{adjustwidth}
		\caption{Applied cyclic load.} 
		\label{fig:paris_load}
	\end{figure}

	The material parameters adopted are $E$ = 6GPa, $\nu = 0.22$, $G_{c} = 2.28$N/mm ($K_\mathrm{IC}$ = 3.69MPa$\sqrt{\mathrm m}$), $\alpha_T=\alpha_N = 9.5\cdot10^{-1}$N/mm$^2$ and $\ell = 0.2$mm. Plane stress conditions are assumed and the simulations are force-controlled. The positive (tensile) cyclic load is characterized by a range $\Delta P$ and minimum, maximum and mean load levels $P_{min}$, $P_{max}$ and $P_{mean}$, respectively (Fig.~\ref{fig:paris_load}).
	
	The tests are performed following the guidelines ASTM E 647 \cite{ASTM_647} for the CT specimen and ASTM E 1820 \cite{ASTM_1820} for the TPB specimen. The crack growth rate curve $d a/d N$ vs. $\Delta K$ is obtained numerically, approximating the crack growth rate for sufficiently small crack length increment $\Delta a$ as 
	
	\eqn{eq:crack_speed}{\frac{d a}{d N} \approx \frac{\Delta a}{\Delta N}\,,}
	
\noindent where a constant value of $\Delta a\simeq$ 0.25 mm is adopted as suggested in \cite{ASTM_647,ASTM_1820}. The stress intensity factor range $\Delta K$ is obtained using the general expression 

\eqn{eq:delta_K}{\Delta K = \frac{\Delta P}{T \sqrt{W}} Y\left( \frac{a}{W} \right)\,,}

\noindent where $W$ and $T$ are respectively a characteristic length and the thickness of the specimen and $Y(a/W)$ is a dimensionless geometric factor whose expression is given in \cite{ASTM_647,ASTM_1820}.	 To obtain the Paris law parameters, the crack growth rate data lying in the Paris regime (i.e., for stable crack propagation) are best fitted to the Paris law to determine $C$ and $m$ in a double logarithmic plot. The algorithms adopted to obtain the numerical crack growth rate curves and the Paris parameters are detailed in \ref{app:paris}.

		\subsubsection{Compact tension (CT) specimen} \label{sct:CT_ex}
\begin{figure}[!h]
	\begin{adjustwidth}{-3cm}{-3cm}
	\centering
		\includegraphics{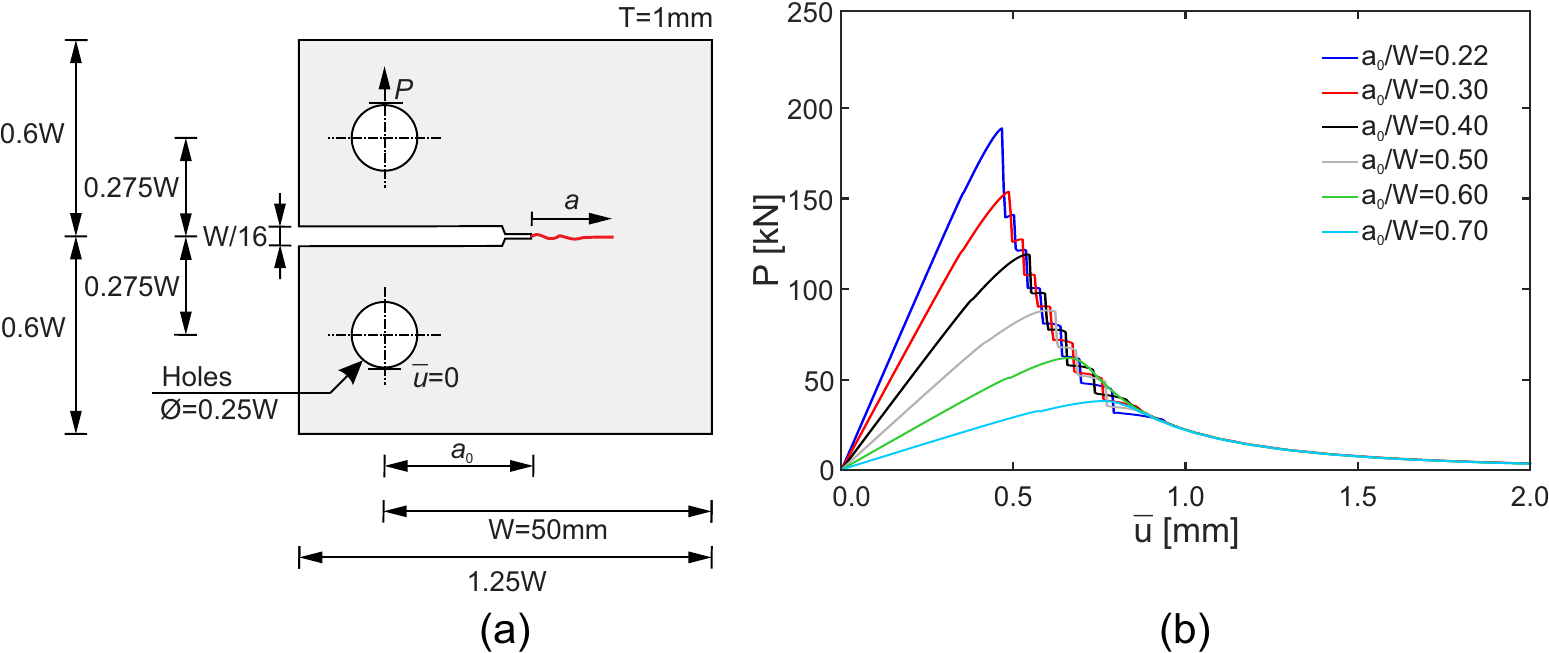}                                                                                             
	\end{adjustwidth}
		\caption{CT specimen: (a) geometry and boundary conditions and (b) monotonic load-displacement curves for different lengths of the initial notch.} 
		\label{fig:CT_geom}
	\end{figure}

		The geometry and boundary conditions used for the CT tests are summarized in Fig.~\ref{fig:CT_geom}a along with the monotonic force-displacement curves for different lengths of the initial notch $a_0$ (Fig.~\ref{fig:CT_geom}b). As for Fig.~\ref{fig:sent_geom}b, also in this case some jumps in the load-displacement curve are observed in the post-peak regime. They are here followed by a limited hardening branch again due to short phases of unstable propagation of the crack, whose tip can reach regions where the fatigue history variable $\fpsi$ is lower than  the threshold value $\alpha_T$ and the fracture toughness is thus still the same of the virgin material. 
		
\begin{figure}[!h]
	\begin{adjustwidth}{-3cm}{-3cm}
	\centering
		\includegraphics{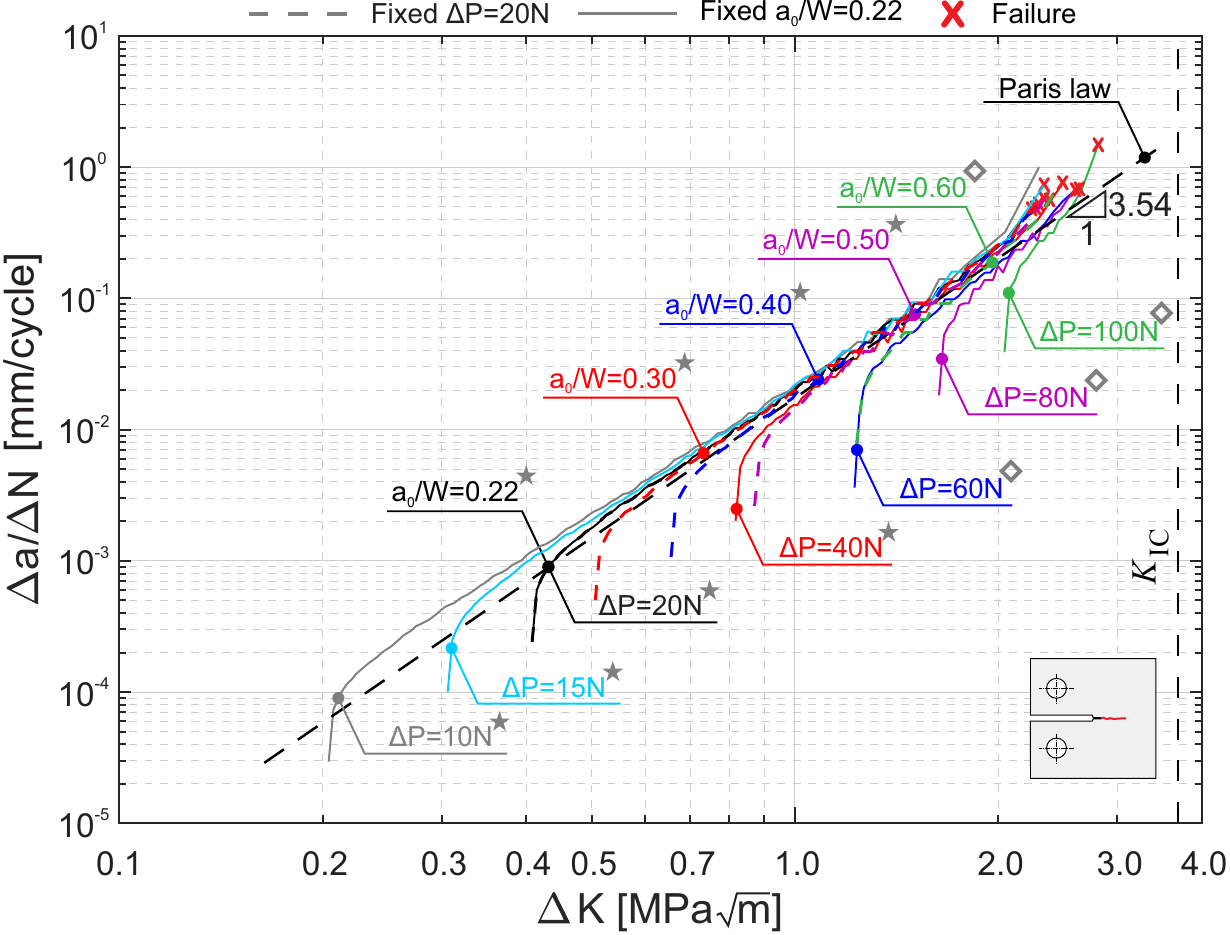}                                                                                             
	\end{adjustwidth}
		\caption{Fatigue crack growth rate curves for the CT specimen varying the length of the initial notch and load range. ($\star$ curves used to obtain the Paris law parameters in Tab.~\ref{tab:Paris_2}; $\diamond$ curves not used to obtain the Paris law parameters because of the absence or scarce extension of the Paris regime.) } 
		\label{fig:CT_paris}
	\end{figure}

For a fixed geometry, the range of the stress intensity factor $\Delta K$ is related only to $\Delta P$ and $a$ (Eq.~\ref{eq:delta_K}). Hence, following the Paris theory, the crack growth rate obtained changing the load range $\Delta P$ or the initial notch $a_0$ should be the same provided $\Delta K$ is the same.  Fig.~\ref{fig:CT_paris} presents the results of different simulations of a cyclic test on the CT specimen varying either $\Delta P$ or $a_0/W$ while keeping the other parameters the same according to Tab.~\ref{tab:Paris_2}.

		\begin{table}
	\begin{adjustwidth}{-3cm}{-3cm}
		\centering
	\begin{tabular}{ccccccc} \toprule
		\multirow{2}{*}{Specimen}& $a_0/W$ &$\Delta P$ & $P_{max}/P_{mon}$& $N_u$& \multirow{2}{*}{$C$} &  \multirow{2}{*}{$m$}\\
		& [-] &[N] & [-]&[-]&   &   \\ \toprule 
		CT & \multirow{ 11}{*}{0.22} & 10.0 &6.7\%{$^\star$}&60974& 2.15$\cdot10^{-2}$  & 3.25 \\
		CT & & 15.0 &10.1\%{$^\star$}&17690& 2.12$\cdot10^{-2}$  & 3.38  \\
		CT & & 20.0 &13.5\%{$^\star$}&7288& 2.01$\cdot10^{-2}$ & 3.42 \\
		CT & & 40.0 & 26.9\%{$^\star$}&817 & 1.60$\cdot10^{-2}$ &  3.73\\
		CT & & 60.0 & 40.4\%{$^\diamond$}&210 & - &- \\
		CT & & 80.0 & 53.8\%{$^\diamond$}& 71 & - &-   \\
		CT & & 100.0 & 67.3\%{$^\diamond$}& 27 & - &- \\ 
		CT & & 110.0 & 74.0\%{$^\diamond$}& 16 & - &- \\ 
		CT & & 120.0 & 80.8\%{$^\diamond$}& 9 & - &-  \\ 
		CT & & 130.0 & 87.5\%{$^\diamond$}& 4 & - &-  \\ 
		CT & & 140.0 & 94.2\%{$^\diamond$}& 1 & - &-  \\ \midrule	
		CT & 0.30 & \multirow{ 4}{*}{20.0} & 16.5\%{$^\star$} & 3645& 1.98$\cdot10^{-2}$& 3.50 \\
		CT & 0.40 & & 21.4\%{$^\star$}  & 1483 & 1.85$\cdot10^{-2}$ & 3.61 \\
		CT & 0.50  &  & 28.9\%{$^\star$}& 512 & 1.57$\cdot10^{-2}$& 3.83\\
		CT  & 0.60 && 41.2\%{$^\diamond$}&129 & -  & -  \\ \midrule
		TPB & \multirow{12}{*}{0.45} & 4.0 & 6.9\%{$^\star$} & 41823& 1.81$\cdot10^{-2}$ &3.25  \\ 	
		TPB & & 5.0 & 8.7\%{$^\star$} & 21198& 1.80$\cdot10^{-2}$  & 3.30 \\ 			
		TPB  & & 7.5 & 13.0\%{$^\star$} &6100 & 1.73$\cdot10^{-2}$  & 3.33\\
		TPB & & 10.0 & 17.3\%{$^\star$}& 2493 & 1.67$\cdot10^{-2}$  & 3.42 \\
		TPB & & 15.0 & 26.0\%{$^\star$}  & 688 & 1.52$\cdot10^{-2}$  & 3.60  \\
		TPB & & 20.0 & 34.6\%{$^\star$}& 266 & 1.29$\cdot10^{-2}$ &3.86 \\
		TPB & & 25.0 & 43.2\%{$^\diamond$} & 123 & - &-   \\
		TPB & & 30.0 & 51.9\%{$^\diamond$}& 62 & -&-   \\ 
		TPB & & 35.0 & 60.6\%{$^\diamond$}& 33 & - &-  \\ 
		TPB & & 40.0 & 69.2\%{$^\diamond$}& 18 & - &-   \\ 
		TPB & & 45.0 & 77.9\%{$^\diamond$}& 9 & - &-   \\
		TPB & & 50.0 & 86.5\%{$^\diamond$}& 4 & - &-   \\ 
		TPB & & 57.5 & 99.5\%{$^\diamond$}& 1 & - &-   \\  \midrule
		\multicolumn{5}{c}{\bf{Average}} &  {\bf{1.78}$\cdot\bf{10^{-2}}$} &{\bf{3.50}} \\
		\toprule 
	\end{tabular}
	\end{adjustwidth}
		\caption{Paris law parameters obtained for the CT and TPB specimens from the curves of Figs.~\ref{fig:CT_paris} and \ref{fig:TPB_paris}.} 
		 \label{tab:Paris_2}

	\end{table}

\begin{figure}[!h]
	\begin{adjustwidth}{-3cm}{-3cm}
	\centering
		\includegraphics{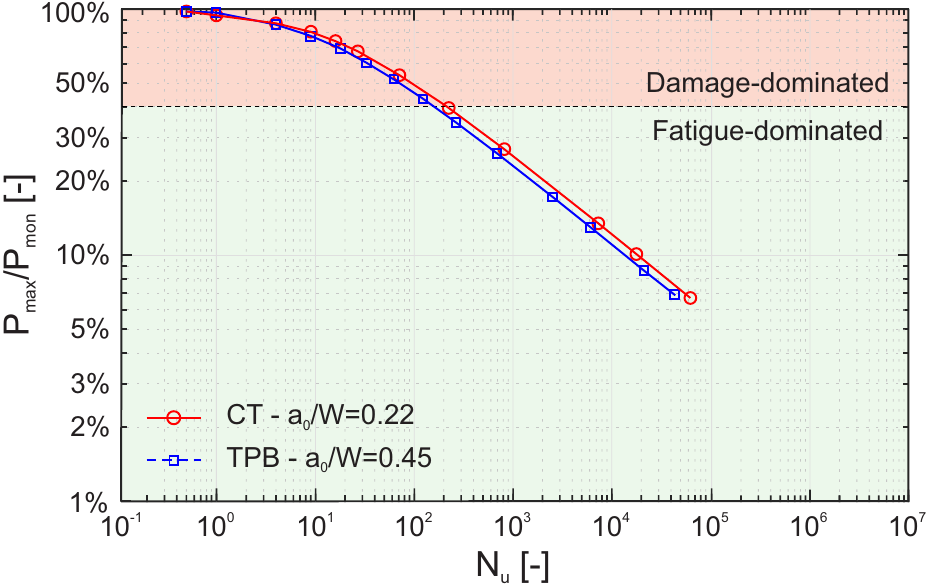}                                                                                             
	\end{adjustwidth}
		\caption{Modified W{\"o}hler curve for the CT and TPB specimens from Tab.~\ref{tab:Paris_2}.} 
		\label{fig:wohler}
	\end{figure}

	Unlike the standard Paris law, the proposed framework is able to recover all the three branches of the experimental crack growth rate curve, featuring an initial short non-linear nucleation phase followed by a linear stable crack propagation branch and, ultimately, by unstable crack propagation (Fig.~\ref{fig:CT_paris}). Also, most of the stable propagation branches of the curves cluster together, with the exception of a few cases (marked with a $\diamond$ in Fig.~\ref{fig:CT_paris}) where the ratio between the maximum load in a cycle $P_{max}$ and the maximum monotonic load $P_{mon}$ is quite high (Tab.~\ref{tab:Paris_2}). For these tests the fatigue life $N_u$ is very limited (i.e., less than 500 cycles), leading to oligocyclic fatigue where the fatigue degradation mechanism cannot develop completely since the damage process dominates. Here the nucleation and unstable crack propagation phases are so close to interfere with each other. A further evidence for a change in the failure mechanism from damage- to fatigue-dominated is given by the modified W{\"o}hler curve relating the ratio $P_{max}/P_{mon}$ to the maximum number of cycles before failure $N_u$, presented in Fig.~\ref{fig:wohler}. For the fatigue model adopted here and based on Fig.~\ref{fig:wohler} the limit between the two regimes is close to $P_{max}/P_{mon}\simeq$40\%.

	  For the curves with $P_{max}/P_{mon}>$40\% that feature a clear stable propagation regime (marked with $\star$ in Fig.~\ref{fig:CT_paris}), it is possible to obtain the Paris law parameters through the procedure illustrated in \ref{app:paris}. The parameters obtained from the different curves are very close (Tab.~\ref{tab:Paris_2}) especially the slope $m$. This confirms that they are material properties which characterize the fatigue behavior, suggesting that the Paris theory is fully recovered by the proposed framework. Note that this result is reached with no fine tuning of the parameters, nor by enforcing a priori the presence of a Paris regime, nor imposing it to be common to all values of $\Delta P$ and $a_0$.	
	  
\begin{figure}[!h]
	\begin{adjustwidth}{-3cm}{-3cm}
	\centering
		\includegraphics{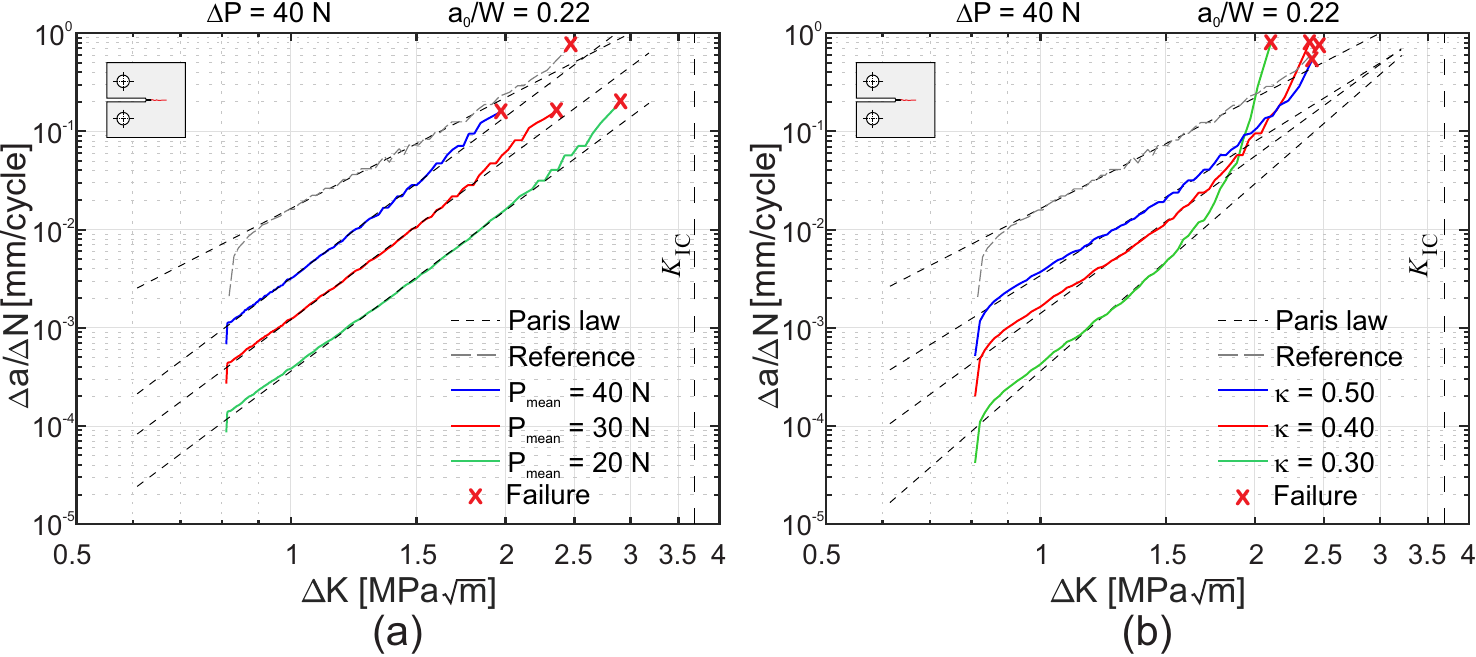}                                                                                             
	\end{adjustwidth}
		\caption{Comparison between a reference fatigue crack growth rate curve obtained using the mean load insensitive accumulation function (Eq.~\ref{eq:no_mean}) and the asymptotic fatigue degradation function (Eq.~\ref{eq:f_basic}) and those obtained adopting (a) the mean load dependent accumulation function (Eq.~\ref{eq:mean}) for different mean load values $P_{mean}$ and (b) the logarithmic fatigue degradation function (Eq.~\ref{eq:f_log}) for different $\kappa$ values.} 
		\label{fig:CT_mean}
	\end{figure}

	  	Fig.~\ref{fig:CT_mean} shows the effects of $P_{mean}$ as given by Eq.~\ref{eq:mean} and of the adoption of the logarithmic fatigue degradation function Eq.~\ref{eq:f_log} on the crack growth rate curves. The curve of Fig.~\ref{fig:CT_paris} for the CT specimen with $a_0/W$=0.22 and $\Delta P$=40N (mean load $P_{mean}$=20N) is used as a reference curve  (dashed gray line in Fig.~\ref{fig:CT_mean}). Adopting Eq.~\ref{eq:mean} and varying only $P_{mean}$ leads to curves whose trend is similar to the reference one but shifted in the $\Delta a  / \Delta N$-$\Delta K$ plane of a quantity dependent on the parameter $\alpha_N$ of Eq.~\ref{eq:mean} (Fig.~\ref{fig:CT_mean}a). Comparing the mean load sensitive curves, one can notice that they are shifted toward higher crack growth rates as the mean load increases. This observation is confirmed by the Paris parameters obtained in Tab.~\ref{tab:Paris_1}, which show negligible differences in $m$ and largely different values of $C$.
		
	\begin{table}
	\begin{adjustwidth}{-3cm}{-3cm}
		\centering
	\begin{tabular}{ccccc} \toprule
		\multirow{2}{*}{Curve} & $\Delta P$ & $a_0/W$ & \multirow{2}{*}{$C$} & \multirow{2}{*}{$m$} \\
			&	[N] & [-] & & \\ \toprule 
		Reference &{40} &{0.22} & 1.60$\cdot10^{-2}$ & 3.73\\ \midrule
		$P_{mean}$=20 N &\multirow{ 3}{*}{40} & \multirow{ 3}{*}{0.22} &  3.56$\cdot10^{-4}$ & 5.42 \\
		$P_{mean}$=30 N  & && 1.19$\cdot10^{-3}$ & 5.39 \\
		$P_{mean}$=40 N  && &  3.15$\cdot10^{-3}$ & 5.48 \\ \midrule
		$\kappa$=0.30    &\multirow{3}{*}{40} & \multirow{ 3}{*}{0.22} &  3.38$\cdot10^{-4}$ & 6.32 \\
		$\kappa$=0.40    &&&  1.32$\cdot10^{-3}$ & 5.30\\
		$\kappa$=0.50    &&&  3.21$\cdot10^{-3}$ & 4.53\\
		\toprule
	\end{tabular}
		\caption{Paris law parameters obtained from the curves of Fig.~\ref{fig:CT_mean}.} 
		 \label{tab:Paris_1}
	\end{adjustwidth}
	\end{table}
	
	Fig.~\ref{fig:CT_mean}b compares the reference curve with the results obtained adopting the logarithmic fatigue degradation function Eq.~\ref{eq:f_log} for different values of $\kappa$. The parameter $\kappa$ permits to control both $C$ and $m$. As highlighted also by Tab.~\ref{tab:Paris_1}, an increasing $\kappa$ leads to higher crack growth rates and lower slopes of the linear branch.

		\subsubsection{Three-point bending specimen} \label{sct:3p_ex}
\begin{figure}[!h]
	\begin{adjustwidth}{-3cm}{-3cm}
	\centering
		\includegraphics{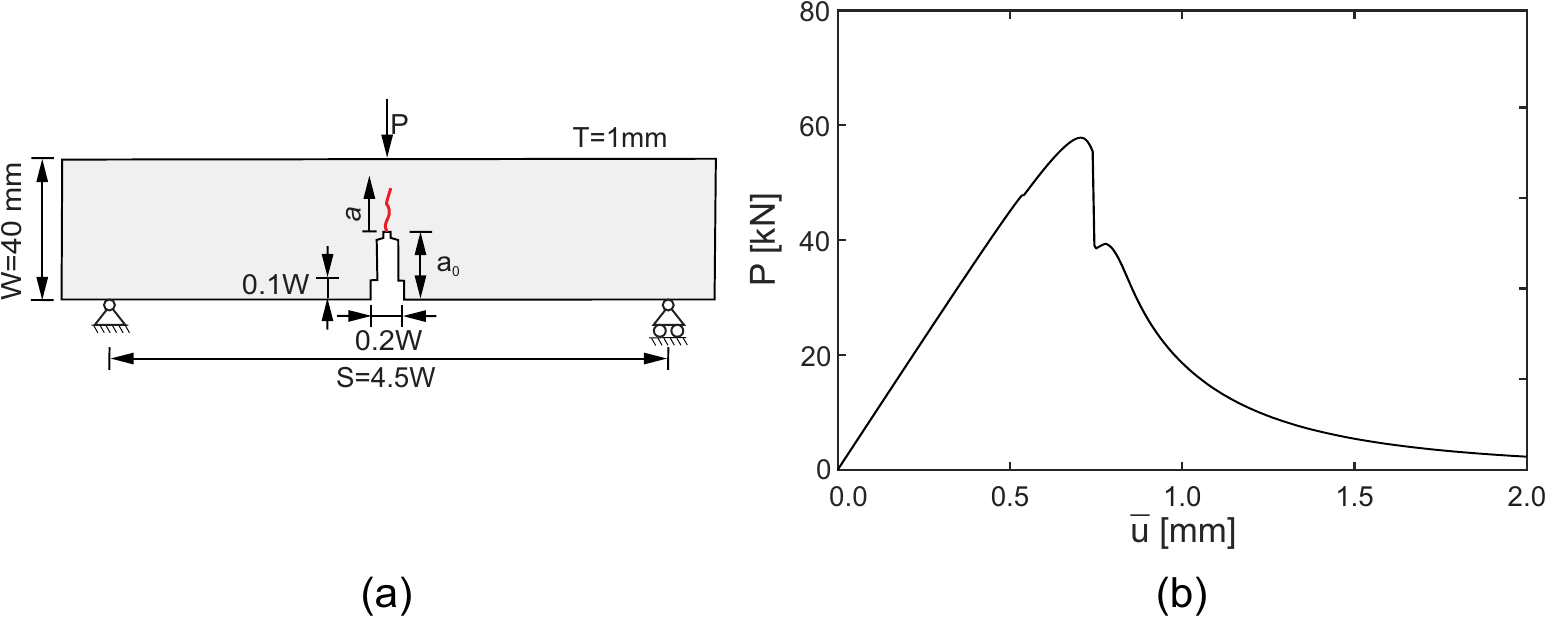}                                                                                             
	\end{adjustwidth}
		\caption{TPB specimen: (a) geometry and boundary conditions and (b) monotonic load-displacement curves.} 
		\label{fig:TPB_geom}
	\end{figure}

		Aim of this section is to compare the results obtained for the CT specimen with those obtained with a TPB specimen, keeping the material parameters unchanged. The geometry of the TPB specimen is illustrated in Fig.~\ref{fig:TPB_geom}a along with the monotonic load-displacement curve (Fig.~\ref{fig:TPB_geom}b), which features a similar post-peak snap-back/local hardening behavior as previously shown for the CT specimen.

\begin{figure}[!h]
	\begin{adjustwidth}{-3cm}{-3cm}
	\centering
		\includegraphics{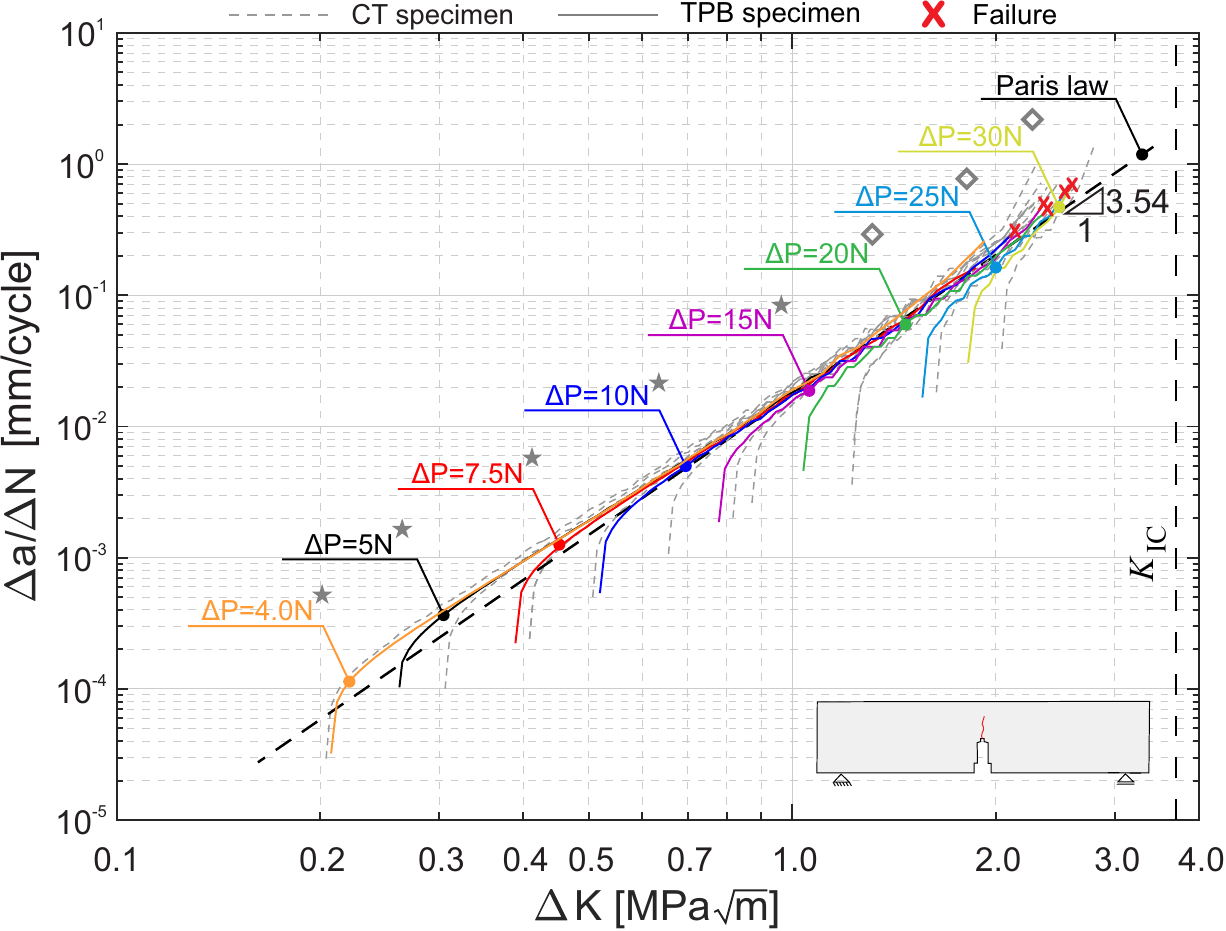}                                                                                             
	\end{adjustwidth}
		\caption{Fatigue crack growth rate curves for the TPB specimen with varying load range and comparison with the CT specimen results. ($\star$ curves used to obtain the Paris law parameters in Tab.~\ref{tab:Paris_2}; $\diamond$ curves not used to obtain the Paris law parameters because of the absence or scarce extension of the Paris regime.)} 
		\label{fig:TPB_paris}
	\end{figure}

 The results obtained in terms of crack growth rate curves for different load ranges $\Delta P$ are illustrated in Fig.~\ref{fig:TPB_paris}, where those obtained for the CT specimen are also reported with gray dashed lines. The results are qualitatively and quantitatively very similar both among the TPB test series and between the latter and those of the CT specimen. The TPB specimen shows as well a transition between damage- and fatigue-dominated behavior (Fig.~\ref{fig:wohler}) with a curve almost identical to the one related to the CT specimens. In particular, the transition takes place for similar $P_{max}/P_{mon}$ ratios, i.e., close to 40\%.		
	
	The Paris parameters for the TPB case are reported in Tab.~\ref{tab:Paris_2}. Again, the values are very similar to each other and to those of the CT test. 
	
	Summing up, the proposed framework reproduces the main features of the Paris theory, including also the nucleation and unstable propagation phases. Moreover, it reproduces a W{\"o}hler-like curve capturing the transition between damage- and fatigue-dominated processes.

	\subsection{Complex geometries and loading conditions} \label{sct:complex_ex}
	After showing that the proposed framework includes and extends the Paris law for brittle materials, we subsequently demonstrate that adopting the phase-field description  of fracture it is possible to deal with complex geometries and loading conditions in 2- and 3-D, which lead to nucleation and propagation of multiple cracks with arbitrarily complex topologies.  
	
		\subsubsection{Plate with holes} \label{sct:plate_holes}
		In this example we study the behavior of a homogeneous plate with 23 uniformly or randomly distributed holes subjected to different loading cycles, including pure compressive and tensile-compressive fatigue. Similar tests, but under monotonic compressive loadings, are investigated experimentally in \cite{Romani2013} and numerically in \cite{Nguyen2016a}. The material parameters are $E =$12 GPa, $\nu = 0.22$, $G_c = 1.40\cdot10^{-3}$ N/mm, $\alpha_T = 6.480\cdot10^{-3}$ N/mm$^2$ and $\ell = 0.018$ mm as in \cite{Nguyen2016a}. Plane strain conditions, displacement control and a uniform spatial discretization with $h = \frac{\ell}{3}$ in the whole domain are assumed.
		
\begin{figure}[!h]
	\begin{adjustwidth}{-3cm}{-3cm}
	\centering
		\includegraphics{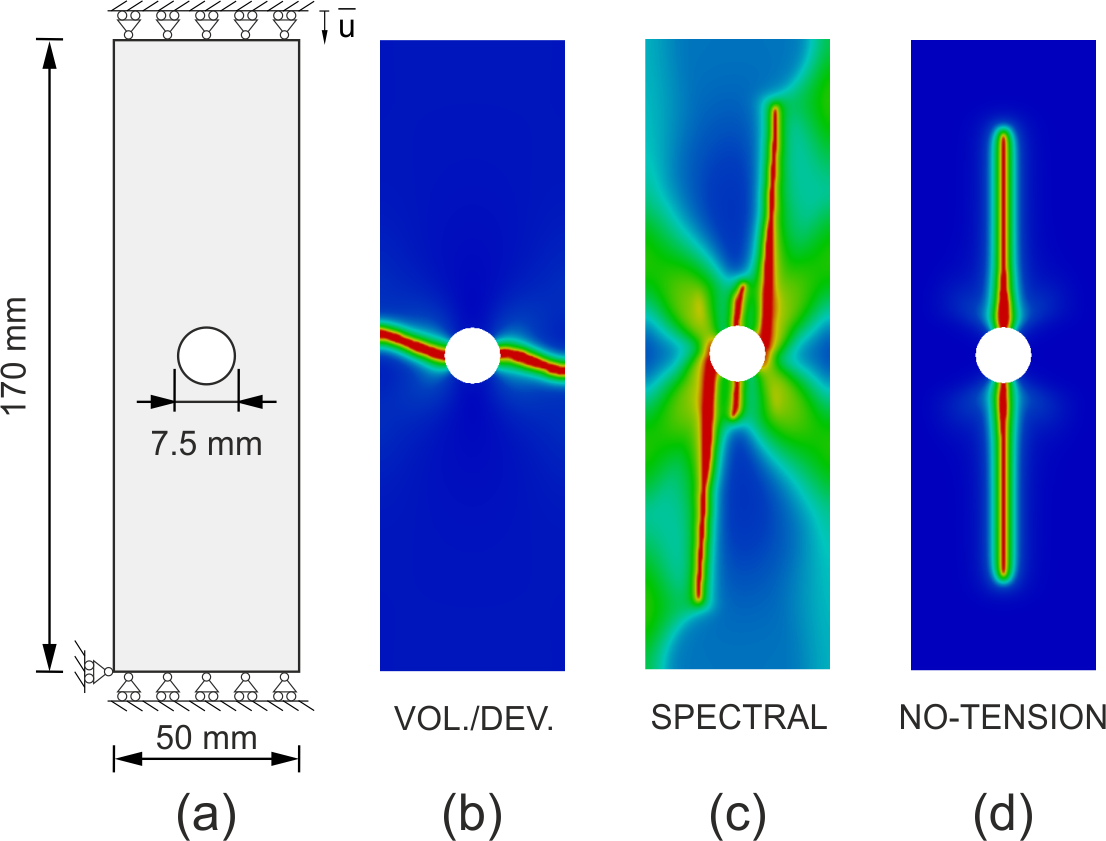}                                                                                             
	\end{adjustwidth}
		\caption{Compression test of a plate with a single hole. (a) Geometry and boundary conditions and results adopting (b) the volumetric/deviatoric split, (c) the spectral split and (d) the no-tension split showing the correct axial splitting failure mechanism. } 
		\label{fig:split_hole}
	\end{figure}

		Before showing results of the fatigue tests on the actual specimens, a premise regarding the choice of the split in compressive tests is needed. When brittle materials are subjected to pure compression, the resulting failure mechanism is characterized by cracks oriented in the direction orthogonal to the principal tensile stress. These cracks nucleate at voids or other (micro-) heterogeneities of the material. This is usually termed axial splitting failure mode \citep{Freddi2010}. To capture this behavior the no-tension split is developed in \citep{Freddi2011}. To show this we simulate a monotonic compression test of a plate with a single hole, whose geometry and boundary conditions are given in Fig.~\ref{fig:split_hole}a. Comparing the resulting crack patterns for the volumetric/deviatoric, spectral and no-tension split (Figs.~\ref{fig:split_hole}b, c and d respectively) it is clear that only the latter reproduces the splitting failure. For this reason, in the fatigue tests under compressive loading described as follows the no-tension split is adopted.
				
\begin{figure}[!h]
	\begin{adjustwidth}{-3cm}{-3cm}
	\centering
		\includegraphics{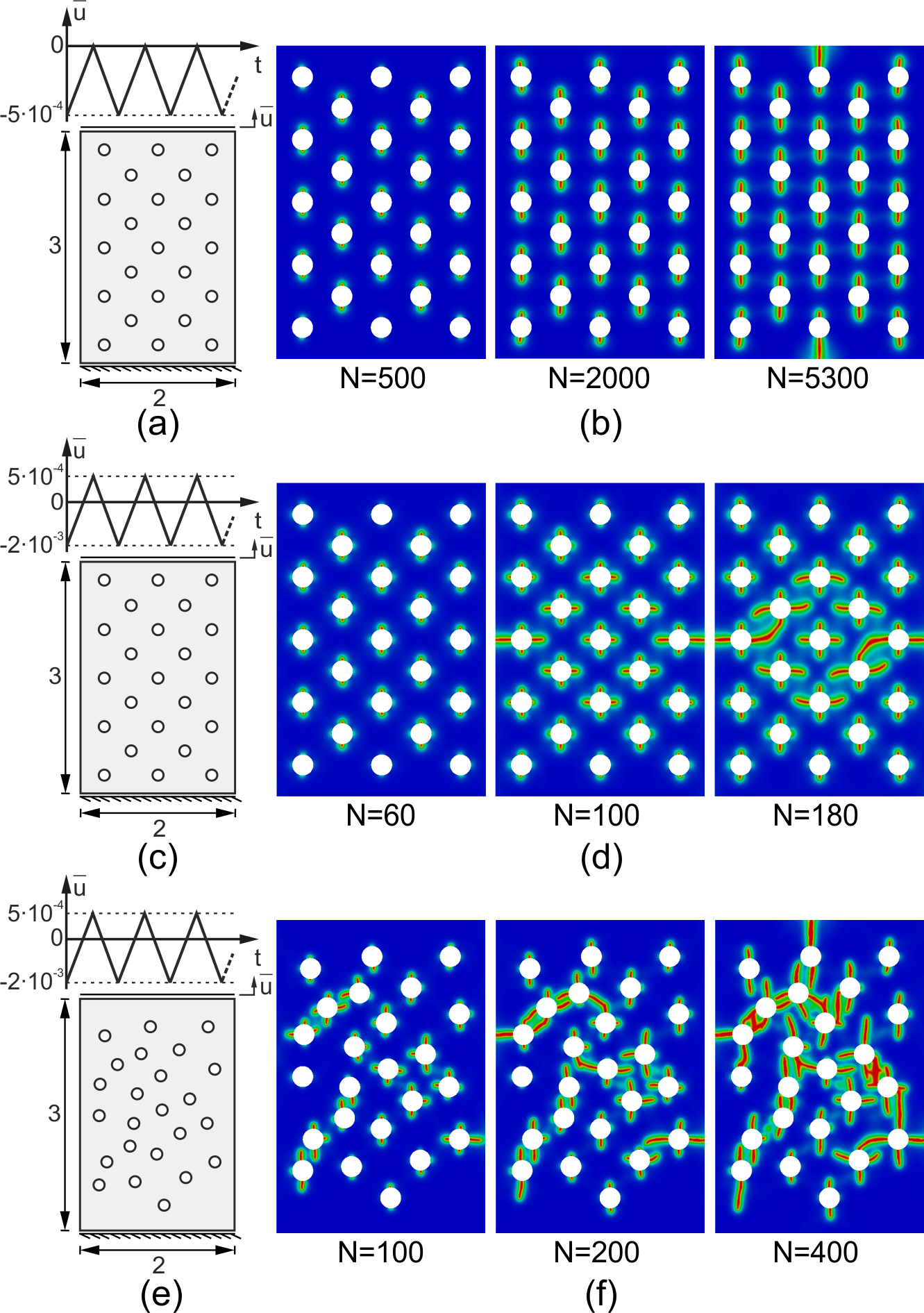}                                                                                             
	\end{adjustwidth}
		\caption{Cyclic test on a plate with 23 holes. Geometry and boundary conditions for (a) plate with uniformly distributed holes and purely compressive load, (c)  plate with uniformly distributed holes and tensile-compressive load and (e)  plate with randomly distributed holes and tensile-compressive load. (b), (d) and (f) show the phase-field contours at crack nucleation (on the left), stable propagation (in the center) and incipient failure (on the right) respectively for the setups (a), (c) and (e).} 
		\label{fig:holes_test}
	\end{figure}

	The first test involves a plate with uniformly distributed holes subjected to a cyclic compressive load as depicted in Fig.~\ref{fig:holes_test}a. The phase-field contours at the nucleation (on the left), stable propagation (in the center) and incipient failure phase (on the right) are presented in Fig.~\ref{fig:holes_test}b, where we can observe that the model is naturally able to handle the presence of different cracks. Moreover, the final crack patterns obtained in the cyclic and monotonic case (the latter is not shown here but very similar to the right contour of Fig.~\ref{fig:holes_test}b) is in good agreement with what observed experimentally in \citep{Romani2013} and numerically in \cite{Nguyen2016a}. The other two tests are related to an alternate tension-compression cyclic load applied to plates with uniformly (Fig.~\ref{fig:holes_test}c) and randomly (Fig.~\ref{fig:holes_test}e) distributed holes. The crack patterns at crack nucleation (on the left), stable propagation (in the center) and right before failure (on the right) are depicted in Figs~\ref{fig:holes_test}d,e, showing that with the proposed approach it is possible to reproduce complex fracture patterns with many cracks interacting with each other, branching and merging until failure of the specimen. Note that the study of these examples  with standard techniques based on the Paris law is rather complicated since it needs the definition of the stress intensity factor range for each crack, which strongly depends on the evolution of the neighboring ones.

		\subsubsection{3-D geometry: steering arm} \label{sct:3d_ex}
\begin{figure}[!h]
	\begin{adjustwidth}{-3cm}{-3cm}
	\centering
		\includegraphics{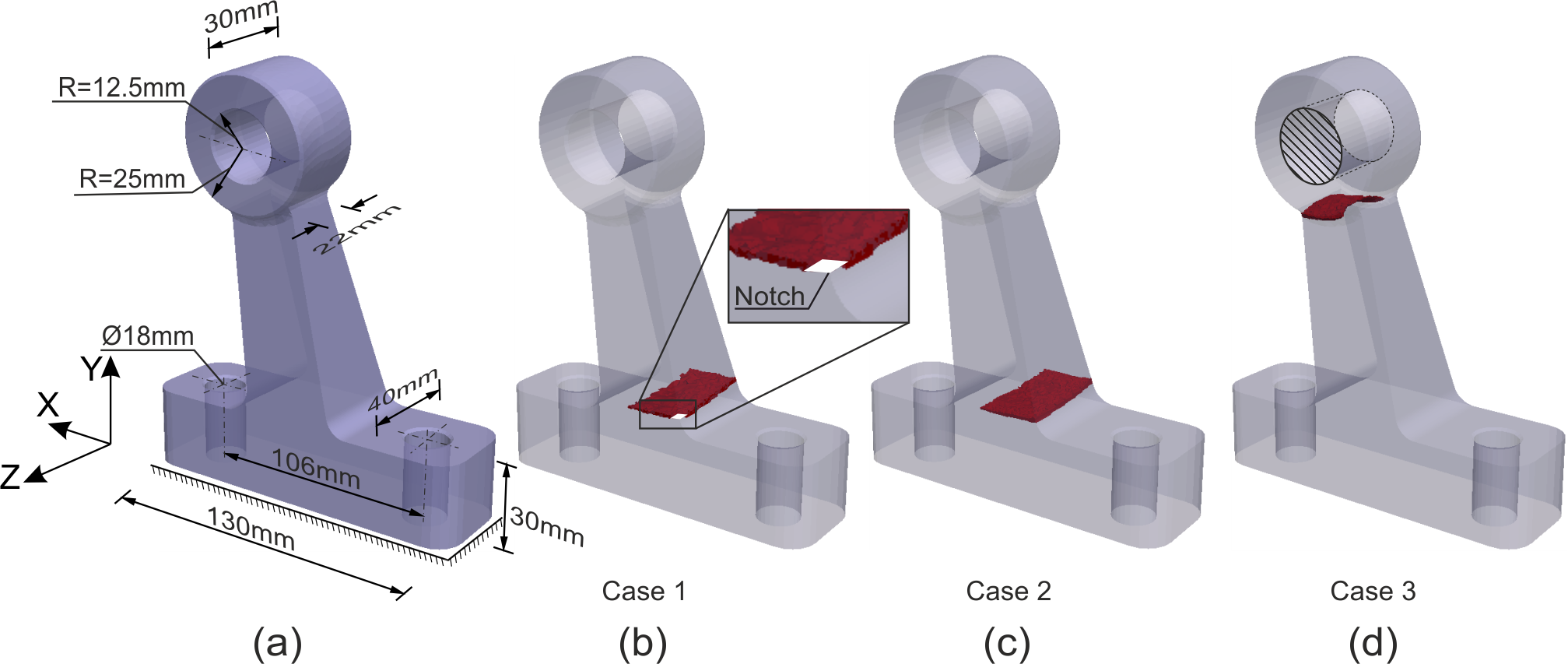}                                                                                             
	\end{adjustwidth}
		\caption{Steering arm: (a) geometry and crack pattern at failure under monotonic loading for (b) \textsf{case1}, (c) \textsf{case 2} and (d) \textsf{case 3}.} 
		\label{fig:3d_mon}
	\end{figure}

		This example aims at investigating the fatigue behavior of a real 3-D component subjected to fatigue. In particular, the steering arm investigated in \citep{Rabold2014} with the geometry illustrated in Fig.~\ref{fig:3d_mon}a is considered. The material parameters adopted are $E =$ 110 GPa, $\nu = 0.30$, $G_c =0.40$ N/mm, $\alpha_T = 4.761\cdot10^{-2}$ N/mm$^2$ and $\ell = 0.7$ mm. Figs.~\ref{fig:3d_mon}b-d illustrate the predicted crack pattern for a monotonic load in positive $x$ direction applied to the internal surface of the upper ring with different boundary conditions. For all the tests the bottom surface of the base and the inner surfaces of the two base holes are fixed in all directions. In \textsf{case 1} (Fig.~\ref{fig:3d_mon}b) there is a notch in the lower part of the shaft as in \citep{Rabold2014}, in \textsf{case 2} there is no notch (Fig.~\ref{fig:3d_mon}c) and \textsf{case 3} (Fig.~\ref{fig:3d_mon}d) is the same as \textsf{case 2} but the inner surface of the upper ring is fixed in the $y$ and $z$ directions. We can see that, if no notch is a priori assumed, the crack can nucleate at both top (\textsf{case 3}) or bottom (\textsf{case 2}) of the shaft depending on the boundary conditions.
			
\begin{figure}[!h]
	\begin{adjustwidth}{-3cm}{-3cm}
	\centering
		\includegraphics{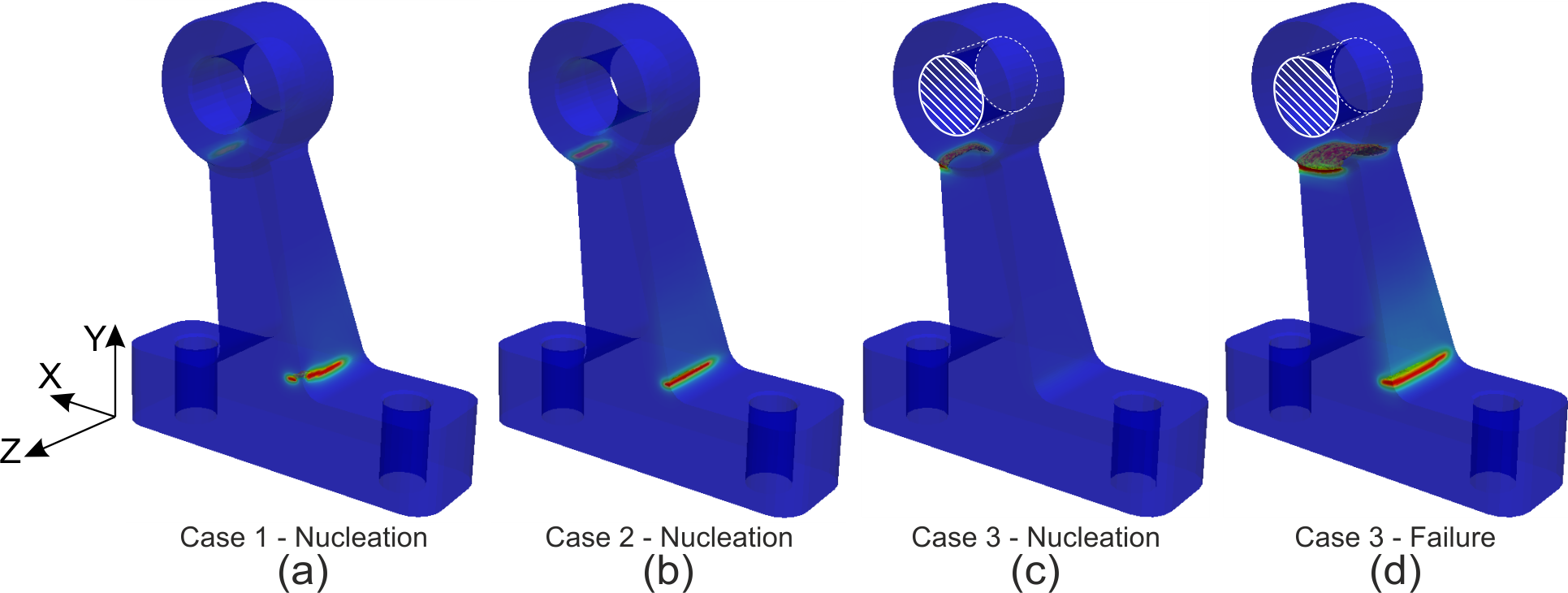}                                                                                             
	\end{adjustwidth}
		\caption{Fatigue test with displacement applied in positive $x$ direction: crack nucleation for (a) \textsf{case1}, (b) \textsf{case 2}, (c) \textsf{case 3} and (d) final crack pattern for  \textsf{case 3}.} 
		\label{fig:3d_cyclic}
	\end{figure}

	Applying a cyclic displacement with range $\Delta \overline u$ = 3.5$\cdot$10$^{-2}$mm in positive $x$ direction the crack nucleates at the location of the monotonic case for all of the three cases analyzed (Fig.~\ref{fig:3d_cyclic}a-c). For the \textsf{case 2} a limited evolution of the phase-field variable, i.e. $d\le$ 0.7, below the upper ring is visible (Fig.~\ref{fig:3d_cyclic}b), which however stops after the nucleation of the main crack in the lower section of the shaft. The final crack pattern for \textsf{case 1} and \textsf{case 2} is very similar to the monotonic case (Figs.~\ref{fig:3d_mon}b-c), while for \textsf{case3} two cracks are present, the one responsible for failure below the upper ring and a secondary one in the lower part of the shaft (Fig.~\ref{fig:3d_mon}d).
				
\begin{figure}[!h]
	\begin{adjustwidth}{-3cm}{-3cm}
	\centering
		\includegraphics{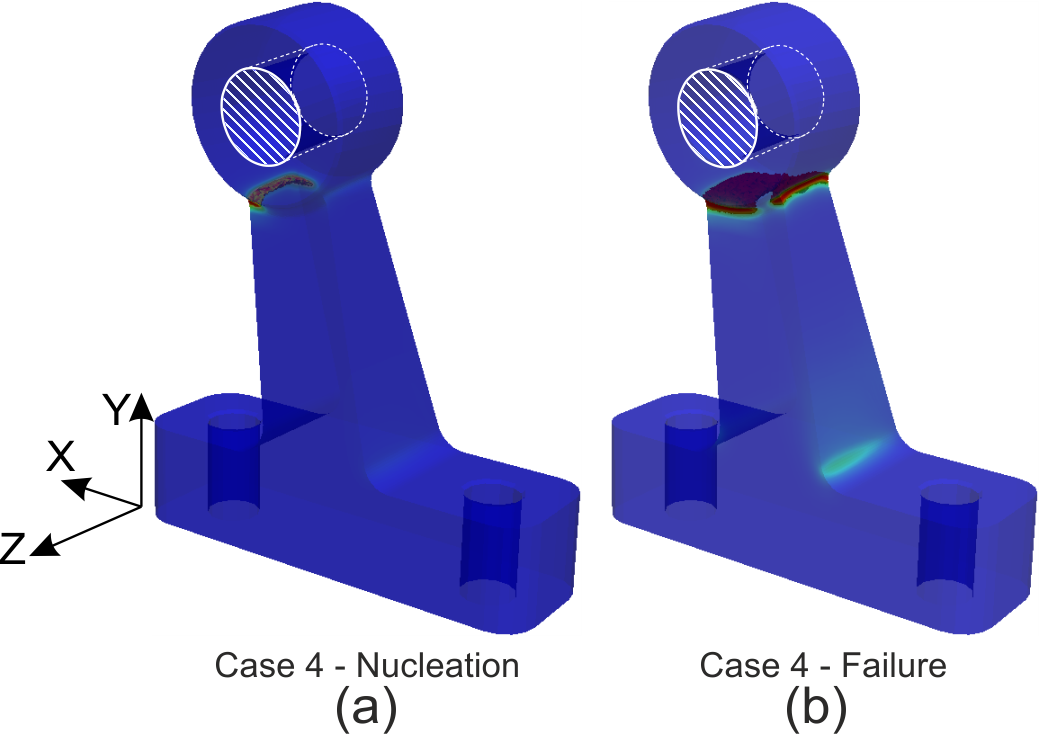}                                                                                             
	\end{adjustwidth}
		\caption{Fatigue test with displacement applied simmetrically in positive and negative $x$ directions (\textsf{case 4}): (a) crack nucleation and (b) final crack pattern.} 
		\label{fig:3d_case4}
	\end{figure}

	A further test, \textsf{case 4}, is performed by applying to \textsf{case 3} a symmetric displacement in positive and negative $x$ directions with range $\Delta \overline u$ = 5$\cdot$10$^{-2}$mm. In this case, the crack pattern becomes more complicated, with two competing cracks nucleating at the top of the shaft that finally cause the failure of the component (Fig.~\ref{fig:3d_case4}). In particular, a crack nucleates first as in \textsf{case 3} (Fig.~\ref{fig:3d_case4}a), while a second one appears later on the opposite side as a result of the negative displacement (Fig.~\ref{fig:3d_case4}b).

	The study of the component of Fig.~\ref{fig:3d_mon}a using standard methods as in \citep{Rabold2014} is possible only assuming an initial notch, which is usually the result of the early stage of the service life. However, imposing a priori shape and position for such a notch in a complex case such as the one studied here is not trivial, being a function of loading and boundary conditions. Conversely, the proposed framework is naturally able to let the fracture nucleate at the most stressed regions.

\section{Conclusions} \label{sct:concl}
A novel framework to model fatigue in brittle materials based on the variational phase-field approach to fracture is proposed. The standard phase-field free energy functional is modified so as to allow the fracture toughness of the material to decrease as a suitable fatigue scalar history variable increases. The reduction rate is governed by a fatigue degradation function that acts as a fatigue constitutive equation and takes as argument only the fatigue history variable. The choice of both history variable and fatigue degradation function is very flexible, being subjected only to some general requirements. In the present work, the fatigue history variable is assumed as the cumulated active part of the elastic strain energy density. Two definitions of the cumulated variable and two fatigue degradation functions are proposed, allowing to reproduce the major fatigue characteristics of different brittle materials. Depending on the specific choice, a maximum of three additional parameters is required.

Based on the obtained results the following conclusions can be drawn:
\begin{itemize}[-]
\item the proposed framework reproduces the Paris theory for fatigue crack growth in brittle materials, encompassing naturally also the nucleation and unstable propagation phases besides the Paris regime; 
\item in addition it enables the computation of the parameters of the Paris law and extensions thereof;
\item the W{\"o}hler or $S-N$ curve and the transition between damage- and fatigue-dominated regimes are also naturally reproduced;
\item the Palmgren-Miner rule and the behavior under monotonic loading are obtained as special cases;
\item the proposed approach is able to naturally handle complex geometries, loading conditions and fracture patterns including multiple cracks with arbitrary topology also in 3-D.
\end{itemize}

\FloatBarrier

\appendix

\section{Relation between the Miner rule and the Paris law}\label{app:miner_appr}
Eq.~\ref{eq:delta_K} can be recasted in the form 

\eqn{eq:sif2}{ \Delta K= y(a) \Delta P \sqrt{\pi a}\,,}

\noindent where $y(a)$ is a modified geometric factor. Substituting Eq.~\ref{eq:sif2} into Eq.~\ref{eq:paris} we have

\eqn{eq:paris2}{\frac{d a}{d N}=C \pi^{m/2}y(a)^m \Delta P^m a^{m/2}\,.}

\noindent Integrating Eq.~\ref{eq:paris2} over a single cycle $N_i$, assuming $a_{i}$ ($a_{i+1}$) is the initial (final) crack length and that  $y(a) = y(a_{i})$ is cycle-wise constant we obtain

\eqn{eq:prop_paris}{a_{i+1}^{1-m/2}-a_{i}^{1-m/2}=\left(1-\frac{m}{2}\right)C \pi^{m/2} y(a_i)^m \Delta P^m =\beta \left(y(a_i) \Delta P \right)^m \,,}

\noindent where $\beta=\left(1-\frac{m}{2}\right)C \pi^{m/2}$ is a constant. 

From Eq.~\ref{eq:prop_paris} it is clear that the crack advancement in a single cycle depends on the current length of the crack. Hence, the contribution of the $i$-th cycle to the crack growth is history dependent. In order to comply with the Palmgren-Miner assumption the right-hand-side of Eq.~\ref{eq:prop_paris} needs to be independent on the current crack length $a_i$, so that the order of the cycles does not influence the crack growth. Such condition can be met cycle-wise only if the geometric factor is constant, i.e. $y(a)=\overline y$ such as in the case of infinite domains \citep{Tada2000}, or changing $\Delta P$ to compensate the variation of $y(a)$ so that $\left(y(a_i) \Delta P\right)^m$ is constant. Since $y(a)$ is generally a monotonic increasing function of $a$, the applied load should decrease with the crack advancement.

\section{Numerical evaluation of the crack growth rate curve}\label{app:paris}
The numerical evaluation of the crack growth rate curve and of the Paris law is performed following the standard guidelines ASTM E 647 \cite{ASTM_647} and ASTM E 1820 \cite{ASTM_1820}. Defining as $a_n$ and $a_{n+1}$ the crack length at the cycle $N_n$ and $N_{n+1}$ respectively, the crack growth rate, which is assumed to remain constant for sufficiently small crack length increments $\Delta a = a_{n+1}-a_n$, is obtained as

\eqn{eq:num_Da}{\left(\frac{\Delta a}{\Delta N}\right)_{n+1/2} = \frac{\Delta a}{N_{n+1}-N_{n}}\,,}

\noindent where $\Delta a \simeq$ 0.25 mm  \cite{ASTM_647,ASTM_1820}. The corresponding stress intensity factor range reads

\eqn{eq:num_DK}{\Delta K_{n+1/2}  = \frac{\Delta P}{T \sqrt{W}} Y_{n+1/2}\left( \gamma_{n+1/2}=\frac{a_{n+1/2}}{W} \right)\,,}

\noindent where $a_{n+1/2} = \frac{1}{2} \left(a_{n+1} + a_n\right)$ is the average crack length and the geometric factor for the CT specimen is  

\eqn{eq:geom_CT}{\begin{split}Y_{n+1/2}\left(\gamma_{n+1/2}\right) = & \frac{(2+\gamma_{n+1/2})}{(1-\gamma_{n+1/2})^{3/2}} \left( 0.886+4.64\gamma_{n+1/2}+\right. \\& \left.-13.32\gamma_{n+1/2}^2+14.72\gamma_{n+1/2}^3-5.6\gamma_{n+1/2}^4 \right)\,,\end{split}}

\noindent while for TPB specimen it is

\eqn{eq:geom_TPB}{\begin{split}Y_{n+1/2} \left(\gamma_{n+1/2}\right) = & \frac{(3 \frac{S}{W} \gamma_{n+1/2})}{2(1+2\gamma_{n+1/2})(1-\gamma_{n+1/2})^{3/2}} \left(1.99-\gamma_{n+1/2}(1-\gamma_{n+1/2})\right. \\& \left. \left( 2.15-3.93\gamma_{n+1/2}+2.7\gamma_{n+1/2}^2 \right) \right)\,.\end{split}}

Once the crack growth rate curve is available, the parameters of the Paris law can be obtained by a linear regression in the double logarithmic plot of the linear stable propagation phase. In this work, the linear regression is performed on the points included in the central third portion of the $\Delta K$ range spanned during the test. The resulting parameters are reported in Tab.~\ref{tab:Paris_2}.


\section*{Acknowledgments}
This work was funded by the DFG through the Research Training Group 2075 "Modelling the constitutive evolution of building materials and structures with respect to aging" and the Priority Program 2020 "Cyclic deterioration of High-Performance Concrete in an experimental-virtual lab" along with the MIUR-DAAD Joint Mobility Program project "Variational approach to fatigue phenomena with phase-field models: modeling, numerics and experiments". Prof. Meinhard Kuna is gratefully acknowledged for providing the 3-D geometry of the steering arm of sect.~\ref{sct:3d_ex}.

\section*{References} \label{ref}


\bibliography{References} 
\bibliographystyle{elsarticle-num2} 

%
\end{document}